\documentclass[a4paper,11pt]{article}
\pdfoutput=1 % if your are submitting a pdflatex (i.e. if you have
\usepackage{jheppub} % for details on the use of the package, please
\usepackage{subfigure}
\usepackage{xcolor}
\usepackage[T1]{fontenc} % if needed
\usepackage{xspace}
\usepackage{multirow}
\usepackage{hyperref}
\usepackage{slashed}
\usepackage{amsmath}

\usepackage[utf8]{inputenc}

\newcommand{\Rambo}{R\protect\scalebox{0.8}{AMBO}\xspace}
\newcommand{\Hone}{H\protect\scalebox{0.8}{1}\xspace}
\newcommand{\Zeus}{Z\protect\scalebox{0.8}{EUS}\xspace}
\newcommand{\HERA}{H\protect\scalebox{0.8}{ERA}\xspace}

\def\GeV{\ifmmode {\mathrm{\ Ge\kern -0.1em V}}\else \textrm{Ge\kern -0.1em V}\fi}%
\def\GeV{\ifmmode {\mathrm{\ Ge\kern -0.1em V}}\else \textrm{Ge\kern -0.1em V}\fi}%

\newcommand{\startappendix}{
\setcounter{section}{0}
\renewcommand{\thesection}{\Alph{section}}
\renewcommand{\theequation}{\Alph{section}.\arabic{equation}}}

\newcommand{\Appendix}[1]{
\refstepcounter{section}
\begin{flushleft}
{\Large\bf Appendix: #1}
\end{flushleft}}

\title{\boldmath Searching for QCD Instantons at Hadron Colliders}

\author{Valentin V. Khoze,}
\author{Daniel L. Milne}
\author{and Michael Spannowsky}

\affiliation{IPPP, Department of Physics, Durham University, Durham DH1 3LE, UK}

\emailAdd{valya.khoze@durham.ac.uk}
\emailAdd{daniel.l.milne@durham.ac.uk}
\emailAdd{michael.spannowsky@durham.ac.uk}

\abstract{
QCD instantons are arguably the best motivated yet unobserved nonperturbative effects predicted by the Standard Model. A discovery 
and detailed study of instanton-generated processes at colliders would provide a new window into the phenomenological exploration of QCD and a vastly
 improved fundamental understanding of its non-perturbative dynamics. Building on the optical theorem, we numerically calculate the total instanton 
 cross-section from the elastic scattering amplitude, also including quantum effects arising from resummed perturbative exchanges between hard gluons
in the initial state, thereby improving in accuracy on previous results. Although QCD instanton processes are predicted to be produced with a large scattering
cross-section at small centre-of-mass partonic energies, discovering them at hadron colliders is a challenging task that requires dedicated search 
strategies. We evaluate the sensitivity of high-luminosity LHC runs, as well as low-luminosity LHC and Tevatron runs. We find that LHC 
low-luminosity runs in particular, which do not suffer from large pileup and trigger thresholds, show a very good sensitivity for discovering QCD 
instanton-generated processes.
}

\begin{document}
\preprint{IPPP/20/44}

\maketitle
\flushbottom
%\newpage

%%%%%%%%%%%%%%%%%%%%%%%%%%%%%%%%%%%%%%%%%%%%%%%

\section{\label{Sec:Intro}Introduction}
\medskip

Instantons are arguably the best motivated non-perturbative effects in the Standard Model (SM), and yet they have not been observed so far. Our motivation in this paper is to re-examine QCD instanton contributions to high-energy scattering processes at hadron colliders building up on the recent work~\cite{Khoze:2019jta}
in establishing a robust QCD instanton computational formalism focussed on applications to proton colliders and to discuss experimental signatures.

The status of the SM as the theory of the currently accessible fundamental interactions in particle physics is well-established. 
To a large extend, the evidence for the SM as the most precise theoretical framework for describing strong and electroweak interactions comes from comparing perturbative calculations with the data from particle experiments. The reliance on the weakly coupled perturbation theory is justified at high energies thanks to the asymptotic freedom in the Yang-Mills theory.
But there is another consequence of the non-Abelian nature of the theory that necessitates an inclusion of non-perturbative effects. The non-Abelian nature of QCD and of the weak interactions is known to
give rise to a rich vacuum structure in the Standard Model.
This vacuum structure is well-understood in the semi-classical picture~\cite{Callan:1976je,Jackiw:1976pf} and amounts to augmenting the perturbative vacuum with an infinite set of topologically non-trivial vacuum sectors in a Yang-Mills theory.

\medskip
Instanton field configurations \cite{Belavin:1975fg}
are classical solutions of Yang-Mills equations of motion in the Euclidean space which interpolate between the different semiclassical vacuum sectors in the theory. At weak coupling instantons provide dominant contributions to the path integral and correspond to quantum tunnelling between different vacuum sectors of the SM. These effects are beyond the reach of ordinary perturbation theory 
and in particular in the electroweak theory they lead to the violation of baryon plus lepton number (B+L), while in QCD instantons processes violate chirality \cite{tHooft:1976snw, tHooft:1976rip},
\begin{equation}
 g + g \, \to\, n_g \times g \,+\, \sum_{f=1}^{N_f} (q_{Rf} +\bar{q}_{Lf})\,,
 \label{eq:inst1}
 \end{equation} 
 where $N_f$ is the number of light (i.e. nearly massless relative to the energy scale probed by the instanton) quark flavours.
The QCD instanton-generated process \eqref{eq:inst1} with two gluons in the initial state going to an arbitrary number of gluons in the final state along with $2N_f$ quarks will be the focus of our discussion in Section~\ref{Sec:Calculation}.

\medskip
The purpose of this paper is to provide the most up-to-date computationally robust calculation of QCD instanton contributions to high-energy scattering processes relevant for hadron colliders. 
At the level of the partonic instanton cross-section, there are two main ingredients in the approach we follow.
We shall use the optical theorem approach that will effectively allow us to sum over all final states with arbitrary number of gluons. This is achieved
by evaluating the imaginary part of the forward elastic scattering amplitude computed in the background of the instanton--anti-instanton configuration.
This formalism was originally developed in \cite{Khoze:1991mx}
based on the instanton--anti-instanton field configuration constructed in \cite{Yung:1987zp}.

The second ingredient of our approach relies on the inclusion of certain higher-order effects in the instanton perturbation theory.
Specifically we will take into account resummed radiative exchanges between the hard partons in the initial state~\cite{Mueller:1990ed,Mueller:1990qa}, as they provide the dominant contribution to breaking the classical scale invariance of QCD in quantum theory. Inclusion of these quantum effects (often referred to in the instanton literature as the hard-hard quantum corrections) is required in order to resolve the well-known non-perturbative infra-red (IR)  problem that arises from contributions of QCD instantons with large scale-sizes, as was first shown in ~\cite{Khoze:2019jta}.
We will see that contributions of QCD instantons with large size
are automatically cut-off by the inclusion of these quantum effects. 

\medskip
To a large extent the theory formalism we employ in this paper for computing QCD instanton rates is the same as in the earlier 
work~\cite{Khoze:2019jta}, but we are able to carry out a more complete evaluation of instanton integrals without relying on the 
saddle-point approximation. Specifically, in Sec.~\ref{sec:master_int} we will numerically compute integrals over all instanton--anti-instanton collective coordinates that correspond to positive modes of the instanton--anti-instanton action. Only the final integration over the single negative mode that gives rise to the imaginary part of the amplitude, as required by the optical theorem, will be carried out in the saddle-point approximation. This provides a more robust prediction leading on average to an order of magnitude increase in instanton partonic cross-sections in our case. There is also a number of other more minor technical improvements, in particular in relation to the computation of the mean number of gluons in the final state in Sec.~\ref{sec:n_gluons}. Our results summarised in Tables~\ref{Table:MainResult} and \ref{Table:HadXSection} present cross-sections for instanton-generated processes at partonic and hadronic levels for the LHC and the Tevatron as well as for 30 TeV and 100 TeV future hadron colliders.

\medskip
In section~\ref{Sec:Recoil} we explain how to generalise the calculation of the instanton process to the case where a jet
is emitted from one of the initial state partons. We find that cross-sections calculated for the processes where the instanton recoils against a jet with large momentum are too small to be observable at any present or envisioned high-energy collider. In order to obtain sensitivity to instantons is to disentangle their spherical radiation profile, made of fairly soft jets from the perturbative backgrounds.

\medskip
The event topology of instanton events with its spherical energy distribution between a large number of final-state objects is visibly distinguishable from the usual few-jets events generated in perturbative-QCD processes at the LHC, as discussed in Sec.~\ref{sec:topology}, but QCD instanton processes occur predominantly at small partonic centre-of-mass energies. The combination of both these characteristics suggests that QCD-instanton events are {\it soft bombs}~\cite{Khoze:2019jta}, using the terminology of Ref.~\cite{Knapen:2016hky}, where
the phenomenology 
of such events was first investigated in the context of beyond the Standard Model physics. In our case the soft bombs are fully Standard Model-made. 
 At high-energy colliders, such events struggle to pass trigger and event reconstruction cuts. In Secs.~\ref{sec:LHC} and \ref{sec:tevatron} we assess whether the comparably large hadronic instanton cross-sections might give rise to visible signatures at hadron colliders, in particular the LHC or the Tevatron.
 Examination of data collected with a minimum bias trigger shows that it should be possible to either discover instantons or severely constrain their cross-section. We conclude with a summary in Sec.~\ref{sec:conclusion}.

\medskip

%%%%%%%%%%%%%%%%%%%%%%%%%%%%%%%%%%%%%%%%%%%%%%%%%%%%
\medskip
\section{\label{Sec:Calculation}Computation of the instanton partonic cross-section}

Instanton gauge fields $A_\mu^{\rm inst} (x)$ are the solutions to the self-duality equation, $F_{\mu\nu}=\tilde{F}_{\mu\nu}$,
and as such, instantons are local minima of the Euclidean action. 
In QCD, the instanton configuration contains the gauge field and the fermion components,
\begin{equation}
A_\mu = A_\mu^{\rm inst} (x)\,, \quad
\bar{q}_{Lf} = \psi^{(0)}(x) \,, \quad q_{Rf} = \psi^{(0)}(x) \,,
\label{eq:inst_sol}
\end{equation}
where the gauge-field $A_\mu^{\rm inst} $ is the BPST instanton solution~\cite{Belavin:1975fg} of topological charge 1,
\begin{eqnarray}
A_\mu^{a\, {\rm inst}}(x) &=& \frac{2 \rho^2}{g} \frac{\bar{\eta}^a_{\mu\nu} (x-x_0)_\nu}{(x-x_0)^2((x-x_0)^2+\rho^2)}\,,
\label{eq:instFT1}
\end{eqnarray} 
and the constants $\bar{\eta}^a_{\mu\nu}$ are the 't Hooft eta symbols \cite{tHooft:1976snw}.
The fermionic components $ \psi^{(0)}$ in \eqref{eq:inst_sol} are known as the instanton fermion zero modes.
They are given by the (non-vanishing) solutions
of the Dirac equation in the $A_\mu^{a\, {\rm inst}}$ instanton background, $\gamma^\mu D_\mu^{ {\rm inst}} \psi^{(0)} \,=0$.
The Euclidean action of the BPST instanton is,
\begin{equation}
S[A_\mu^{\rm inst}] =S_I= \frac{8\pi^2}{g^2} = \frac{2\pi}{\alpha_s(\mu_r)} \,,
\label{eq:SI}
\end{equation}
where for the later convenience we have included the dependence of the coupling constant on the RG scale $\mu_r$.
For more detail on instantons and their applications relevant to the material in this section, an interested reader can consult  
a selection of review articles in Refs.~\cite{Coleman:1985rnk,Vainshtein:1981wh,Mattis:1991bj,Schafer:1996wv,Dorey:2002ik}.
Our presentation in sections~\ref{sec:2.1} and \ref{sec:2.2} follows a recent overview of QCD instanton calculus in Ref.~\cite{Khoze:2019jta}.

%%%%%%%%%%%%%%%%%%%%%%%%%%%%%%%%%%%%%%%%%%%%%%%%%%%%
\subsection{QCD instantons and scattering amplitudes}
\label{sec:2.1}
%%%%%%%%%%%%%%%%%%%%%%%%%%%%%%%%%%%%%%%%%%%%%%%%%%%%

The scattering amplitude for the $2 \to n_g+2N_f$ instanton-generated process 
\eqref{eq:inst1} is computed by expanding the path integral around the instanton field configuration \eqref{eq:inst_sol}.

\medskip
\noindent The amplitude
takes the form of an integral over the instanton collective coordinates,
\begin{eqnarray} 
{\cal A}_{\, 2\to\, n_g+ 2N_f} &=&\int d^4 x_0 \int_0^\infty d\rho \, D(\rho) \, e^{-S_I}\,
\prod_{i=1}^{n_g+2} A_{{\rm LSZ}}^{\rm inst}(p_i;\rho)\, \prod_{j=1}^{2N_f} \psi^{(0)}_{{\rm LSZ}}(p_j; \rho)
 \,.
\label{eq:ngA1}
\end{eqnarray}
The integral \eqref{eq:ngA1} is over the instanton position $x_0^\mu$ and the scale-size collective coordinate $\rho$, and it involves the instanton density function $D(\rho)$, 
the semiclassical suppression factor $ e^{-S_I}$ by the instanton action \eqref{eq:SI}, and the product of vector boson 
and fermion field configurations, one for
each external leg of the amplitude, computed on the instanton solutions, and LSZ-reduced.

The instanton density $D(\rho)$ in  \eqref{eq:ngA1} 
arises from computing quadratic fluctuation determinants in the instanton background in the path integral. This is a one-loop effect in the perturbation theory around the instanton and the result is given by~\cite{tHooft:1976snw} ,
\begin{equation}
D(\rho,\mu_r)\,=\, 
\kappa \, \frac{1}{\rho^5} \left( \frac{2\pi}{\alpha_s(\mu_r)}\right)^{2N_c}\, (\rho \mu_r)^{b_0}\,,
\label{eq:Imeasure}
\end{equation}
where $\kappa$ is the normalisation constant of the instanton density in the $\overline{\rm MS}$ 
scheme \cite{tHooft:1986ooh,Hasenfratz:1981tw,Luscher:1981zf}, 
 \begin{equation}
 \kappa\,=\, \frac{2\, e^{5/6-1.511374 N_c}}{\pi^2 (N_c-1)!(N_c-2)!} \, e^{0.291746 N_f} 
\,\simeq\,0.0025 \, e^{0.291746 N_f} \,, 
 \label{eq:kapdef}
 \end{equation}
and $b_0=(11/3)N_c-(2/3) N_f$. 

\medskip
Expressions for the LHZ-reduced instanton field insertions on the right hand side of the integral in \eqref{eq:ngA1} are obtained from
the momentum-space representation of the instanton solution \eqref{eq:instFT1},
\begin{eqnarray}
A^{a\, {\rm inst}}_{LSZ}(p,\lambda) \,=\, \lim_{p^2\to 0} p^2 \epsilon^\mu(\lambda) \, A_\mu^{a\, {\rm inst}}(p)
 \,=\, \epsilon^\mu(\lambda) \,\bar{\eta}^a_{\mu\nu} p_\nu \, \frac{4i\pi^2 \rho^2}{g} \, e^{ip \cdot x_0}
 \,, \label{eq:Inst_LSZA1}
\end{eqnarray}   
where $\epsilon^\mu(\lambda)$ is the polarisation vector for a gluon with a helicity $\lambda$. A similar expression also holds for the LSZ-amputated fermion zero modes, in this case, $\psi^{(0)}_{{\rm LSZ}} \propto \rho$ rather than 
$A^{ {\rm inst}}_{LSZ} \propto \rho^2$ for the gauge field.

\medskip

Combining all the ingredients above, it is now easy to see that the $\rho$-integral in the leading-order instanton amplitude \eqref{eq:ngA1} is power-like divergent --  a well-known fact that signals the breakdown of the leading-order instanton calculation in QCD at large distances ($\rho \gtrsim 1/\Lambda$) where the 
coupling becomes strong and the semi-classical approximation is invalidated.
Instantons are solutions to classical equations and unless quantum effects due to field fluctuations around instantons are appropriately taken into account,
 there is no scale in the microscopic QCD Lagrangian to cut-off large values of the instanton size -- $\rho$ is a classically flat direction.
 To break classical scale-invariance we need to 
 include quantum corrections that describe interactions of the external states. 
 This   
 amounts to inserting propagators in the instanton background between pairs of external fields in the pre-exponential factor 
 in \eqref{eq:instFT1}  and re-summing the resulting perturbation theory.  The dominant effect comes from 
 interactions between 
 the two initial hard gluons \cite{Mueller:1990ed} (these are the states that carry the largest kinematic invariant  $p_1\cdot p_2 = \hat{s}/2$).
 In Ref.~\cite{Mueller:1990qa} Mueller has shown that these quantum corrections formally exponentiate and the resulting expression for 
 the resummed quantum corrections around the instanton 
 generates the factor,
 \begin{eqnarray}
 e^{- (\alpha_s(\mu_r)/16 \pi)  \, \rho^2 E^2 \log E^2/\mu_r^2}\,,
 \label{eq:Mueller_corr}
\end{eqnarray}   
where $E$ is the partonic CoM energy, $E^2 \equiv \hat{s}$. 
This exponential factor
provides an automatic cut-off of the large instanton sizes and the instanton integral over $\rho$ 
can now be safely evaluated.

\medskip

To proceed,
we need to select a value the renormalisation scale $\mu_r$.
Recall that the integrand in \eqref{eq:instFT1} contains the factor,
\begin{equation}
(\rho \mu_r)^{b_0}\,
e^{-\,\frac{2\pi}{\alpha_s(\mu_r)}} \,=\, e^{-\,\frac{2\pi}{\alpha_s(1/\rho)}}\,,
\label{eq:RGinv1}
\end{equation}
where $(\rho \mu_r)^{b_0}$ comes from the instanton density and 
the factor $e^{-\,\frac{2\pi}{\alpha_s(\mu_r)}}$ accounts for the contribution of the instanton action $S_I = \frac{2\pi}{\alpha_s(\mu_r)}$.
The r.h.s. of \eqref{eq:RGinv} is RG-invariant at one-loop, it does not depend on the choice of $\mu_r$, instead the scale of the running coupling constant is set 
at the inverse instanton size. To take advantage of this and to remove large powers of $\rho$ from the integrand, from now on and until the end of this section, we will set the RG scale value at the instanton size,
 \begin{equation}
 \mu_r \,=\, 1/\rho\,.
 \label{eq:RGinv}
\end{equation}

\noindent The amplitude integrand including the Mueller's exponentiated quantum effect is given by,
\begin{eqnarray} 
{\cal A}_{\, 2\to\, n_g+ 2N_f} &=& \kappa \int d^4 x_0\,  \int_0^\infty \frac{d\rho}{\rho^5}   \left( \frac{2\pi}{\alpha_s}\right)^6\,
 e^{-\,\frac{2\pi}{\alpha_s(1/\rho)}\,-\, \frac{\alpha_s(1/\rho)}{16\pi}  \rho^2 E^2 \log E^2\rho^2   }
\nonumber \\
&&\qquad \qquad \qquad \qquad \qquad  \times\, \prod_{i=1}^{n_g+2} A_{{\rm LSZ}}^{\rm inst}(p_i;\rho)\, \prod_{j=1}^{2N_f} \psi^{(0)}_{{\rm LSZ}}(p_j; \rho)
\label{eq:ngA1Q}
\end{eqnarray}

Keeping a careful track of the powers of $\rho$,
the resulting integral in~\eqref{eq:ngA1Q} is proportional to the following expression 
(we note that the integral over the instanton position $\int d^4 x_0$ gives the delta function of the momentum conservation which we drop, along with the overall constant and $\rho$-independent factors),
\begin{eqnarray} 
{\cal A}_{\, 2\to\, n_g+ 2N_f}  &\sim&  \int_0^\infty d\rho\,\, (\rho^2)^{n_g+2 + N_f-5/2}\, e^{ - \frac{\alpha_s(1/\rho)}{16\pi} E^2\rho^2 \log (E^2\rho^2)\,-\, 
\frac{2\pi}{\alpha_s(1/\rho)}}
 \,. 
\label{eq:ngA}
\end{eqnarray}
The integral is no longer divergent in the IR limit of large $\rho$ and can be evaluated and the resulting expression for the amplitude can be used to compute the instanton cross-section. 
In the following section we will obtain the instanton cross-section 
in a more efficient manner using the Optical theorem approach in the following section (Sec.~\ref{sec:2.2}).

\medskip

Before we conclude this section, we would like to comment on the structure of the 
leading-order instanton expression \eqref{eq:ngA1Q}.
Note that the integrand on the right hand side of \eqref{eq:ngA1Q} contains a simple product of bosonic and fermionic components of instanton field configurations, one for each external line of the amplitude.
Such fully factorised structure
of the field insertions implies that at the leading order in instanton perturbation theory
 there are no correlations between the momenta of the external 
legs in the instanton amplitude. Emission of individual particles in the final state are mutually independent, apart from the overall momentum conservation.
The expression in \eqref{eq:ngA1Q} looks like a multi-particle point-like vertex integrated over the instanton position and size.
Thanks to its point-like structure, the instanton vertex in the centre of mass frame describes the scattering process into 
a spherically symmetric multi-particle final state.  The number of gluons $n_g$ is unconstrained and can be as large 
as is energetically viable~\cite{Ringwald:1989ee,Arnold:1987mh} (in practice, 
the dominant contribution will come from $\langle n_g\rangle \sim 4\pi/\alpha_s \gg 1$), 
and a fixed number of quarks (a $q_{L} \bar{q}_{R}$ pair for each light quark flavour).

\bigskip
%%%%%%%%%%%%%%%%%%%%%%%%%%%%%%%%%%%%%%%%%%%%%%%%%%%%
\subsection{The Optical theorem approach}
\label{sec:2.2}
%%%%%%%%%%%%%%%%%%%%%%%%%%%%%%%%%%%%%%%%%%%%%%%%%%%%
\medskip

To compute a total parton-level instanton cross-section $\hat\sigma_{\rm tot}^{\rm inst}$
for the process $gg \to X$,  we use the optical theorem  to relate the cross-section to the imaginary part of the
forward elastic scattering amplitude computed in the background of the instanton--anti-instanton ($I\bar{I}$) configuration,
\begin{eqnarray}
\hat\sigma_{\rm tot}^{\rm inst} &=&  \frac{1}{E^2}\, {\rm Im} \, {\cal A}^{{I\bar{I}}}_4 (p_1,p_2,-p_1,-p_2) 
\,,
\label{eq:op_th_0}
\end{eqnarray}
where $E=\sqrt{\hat{s}} = \sqrt{(p_1+p_2)^2}$ is the partonic CoM energy.

\medskip

For reader's convenience in Appendix~A we outline main steps of the formalism to represent the forward elastic scattering amplitude 
as the integral over 
collective coordinates of the instanton--anti-instanton field configuration following the valley method approach 
developed in \cite{Balitsky:1986qn,Yung:1987zp,Khoze:1991mx,Khoze:1991sa,Verbaarschot:1991sq,Balitsky:1992vs}.  

For our purposes it is sufficient to simply note that the instanton--anti-instanton gauge field is a trajectory in the topological charge zero sector
of the field configuration space parameterised by instanton and anti-instanton collective coordinates.
This trajectory interpolates between the sum of infinitely separated instanton and anti-instanton  and the perturbative vacuum,
\begin{eqnarray}
R\to \infty \,: \qquad &&A_{\mu}^{I\bar{I}}(x) \, \longrightarrow\, A_{\mu}^{I}(x-x_0) \,+\, A_{\mu}^{\bar{I}}(x-x_0-R)\,,\\
R\to 0 \,: \qquad &&A_{\mu}^{I\bar{I}}(x) \, \longrightarrow\, 0\,.
\end{eqnarray}
The configuration $A_{\mu}^{I\bar{I}}(x)$ 
 for arbitrary values of the collective coordinates is determined by solving the gradient flow equation known as the valley equation.
 
The collective-coordinate integral for the amplitude reads,
\begin{eqnarray}
{\cal A}^{{I\bar{I}}}_4 (p_1,p_2,-p_1,-p_2) 
&= &   \int_0^\infty d\rho \int_0^\infty d\bar{\rho}\int d^4R \int d\Omega
D(\rho) D(\bar{\rho}) \,
e^{-S_{I\bar{I}} -\frac{\alpha_s}{16\pi}(\rho^2+\bar{\rho}^2)  E^2 \log \frac{E^2}{\mu_r^2}
} \nonumber  \\
&& \, A^{\rm inst}_{LSZ}(p_1)\,A^{\rm inst}_{LSZ}(p_2)\,A^{\overline{\rm inst}}_{LSZ}(-p_1)\,A^{\overline{\rm inst}}_{LSZ}(-p_2)
\,\, {\cal K}_{\rm ferm} 
\,.
\label{eq:op_th}
\end{eqnarray}
In the expression above we integrate over all collective coordinates:
 $\rho$ and $\bar{\rho}$ are the instanton and anti-instanton sizes,
$R_\mu=(R_0,\vec{R})$  is the separation between the $I$ and $\bar{I}$  positions in the Euclidean space
and $\Omega$ is the $3\times 3$ matrix of relative ${I\bar{I}}$ orientations in the $SU(3)$ colour space.
$D(\rho)$ and  $D(\bar{\rho})$ represent the instanton and the anti-instanton densities \eqref{eq:Imeasure} and 
the field insertions $A^{\rm inst}_{LSZ}(p)$ and $A^{\overline{\rm inst}}_{LSZ}(p')$ are the LSZ-reduced instanton and anti-instanton fields
\eqref{eq:Inst_LSZA1}.
For each pair of the gluon legs with the same incoming/outgoing momentum we have,
 \begin{equation}
 \frac{1}{3}\sum_{a=1}^3 \, \frac{1}{2} \sum_{\lambda=1,2} \, A^{a\, {\rm inst}}_{LSZ}(p,\lambda)\, A_{LSZ}^{a\, {\overline{\rm inst}}}(-p;\lambda) \, =\,
 \frac{1}{6}\, 
 \left(\frac{2\pi^2}{g} \rho \bar\rho \,\sqrt{s'}\right)^2\, e^{iR\cdot p} \,,
\label{eq:LSZA1}
\end{equation}   
and now for the combination of all four external gluons insertions in \eqref{eq:op_th} we have,
\begin{equation}
 A^{\rm inst}_{LSZ}(p_1)\,A^{\rm inst}_{LSZ}(p_2)\,A^{\overline{\rm inst}}_{LSZ}(-p_1)\,A^{\overline{\rm inst}}_{LSZ}(-p_2) \,=\,
 \frac{1}{36}\, \left(\frac{2\pi^2}{g} \rho \bar\rho\, \sqrt{s'}\right)^4\, e^{iR\cdot (p_1+p_2)} \,.
\label{eq:LSZA}
\end{equation}    
The contribution $e^{iR\cdot (p_1+p_2)}$ arises from the
exponential factors $e^{ip_i\cdot x_0}$ and $e^{-ip_i\cdot \bar{x}_0}$ from the two instanton and two anti-instanton legs, which upon the Wick rotation to the 
Minkowski space becomes $e^{R_0 \sqrt{s'}}$. 

The final factor on the right hand side of \eqref{eq:op_th} (apart from the expression in the exponent) is the overlap of fermion zero modes
${\cal K}_{\rm ferm}$ which we will define near the end of the section.

We now turn to the exponent in \eqref{eq:op_th}.
The action of the instanton--anti-instanton configuration was computed
 in \cite{Khoze:1991mx,Khoze:1991sa,Verbaarschot:1991sq}, it is a function of a single 
 variable $z$ known as 
 the conformal ratio of the (anti)-instanton collective coordinates,
\begin{equation}
z\,=\, \frac{R^2+\rho^2+\bar{\rho}^2+\sqrt{(R^2+\rho^2+\bar{\rho}^2)^2-4\rho^2\bar{\rho}^2}}{2\rho\bar{\rho}}\,.
\label{eq:zdef}
\end{equation}
 and takes the form $S_{I\bar{I}}(z)= \frac{4\pi}{\alpha_s} \, {\cal S}(z)$ where,
\begin{eqnarray}
{\cal S}(z) \,=\, 3\frac{6z^2-14}{(z-1/z)^2}\,-\, 17\,-\, 
3 \log(z) \left( \frac{(z-5/z)(z+1/z)^2}{(z-1/z)^3}-1\right)\,.
 \label{eq:Szdef}
\end{eqnarray}
 For more detail on the derivation of the instanton--anti-instanton valley trajectory and the plot of the action as the function of the inter-instanton 
 separation we refer the reader to Appendix~A and Refs.~\cite{Yung:1987zp,Khoze:1991mx,Khoze:1991sa,Verbaarschot:1991sq}.
 
 The second term in the exponent in \eqref{eq:op_th} is recognised as the Mueller's quantum effect 
 of the hard-hard gluon exchanges in the initial state \eqref{eq:Mueller_corr}
 and the similar factor for the anti-instanton gluon exchanges in the final state. 
 
 \medskip
 
The final factor appearing in \eqref{eq:op_th} that needs to be defined, is ${\cal K}_{\rm ferm} (z)$. 
This simply comes from calculating the overlap between the instanton and anti-instanton fermion zero modes \citep{Shuryak:1991pn},
\begin{equation}
\omega=\int d^{4}x \psi_{0}^{\bar{I}}\left(x\right)i\slashed{D}\psi_{0}^{I}\left(x\right).
\end{equation}
\citep{Shuryak:1991pn} also found an integral expression for this which was then calculated analytically in \citep{Ringwald:1998ek} and this expression is then raised to the power $2N_{f}$, the number of fermions.
It arises from the $2N_f$ fermions in the final state of the process \eqref{eq:inst1}. As the instanton--anti-instanton action function ${\cal S}(z)$, the 
fermion factor ${\cal K}_{\rm ferm} (z)$ is a function of a single variable -- the conformal ratio $z$ defined in \eqref{eq:zdef}.
We have,
\begin{equation}
{\cal K}_{\rm ferm} \,=\, (\omega_{\rm\,  ferm})^{2N_f}\,, 
\label{eq:Kferm}
\end{equation}
where $\omega_{\rm\,  ferm} (z)$ was computed in \cite{Ringwald:1998ek},
\begin{equation} 
\omega_{\rm\,  ferm}(z) \,= \,  
\frac{3\pi}{8}\frac{1}{z^{3/2}}\,\,{}_2F_{1} \left(\frac{3}{2},\frac{3}{2};4;1-\frac{1}{z^2}\right)
\,.
\label{eq:omegaF}
\end{equation}

\bigskip 
Putting everything together we can now write down the instanton cross-section \eqref{eq:op_th_0}
 as the finite-dimensional integral in the form,
 \begin{eqnarray}
\label{eq:firstint}
\hat\sigma_{\rm tot}^{\rm inst} &\simeq& 
\frac{1}{E^2}\,{\rm Im}\, \frac{\kappa^2 \pi^4}{36\cdot4}   \int \frac{d\rho}{\rho^5}
 \int \frac{d\bar{\rho}}{\bar{\rho}^5}  \int d^4 R \int d\Omega  \left( \frac{2\pi}{\alpha_s(\mu_r)}\right)^{14} (\rho^2 E)^2 (\bar{\rho}^2E)^2\,
  {\cal K}_{\rm ferm}(z)
% \times 
\nonumber
  \\
&& \,
 (\rho \mu_r)^{b_0} (\bar{\rho} \mu_r)^{b_0}\,\exp\left(R_0 E\,-\,\frac{4\pi}{\alpha_s(\mu_r)} \,{\cal S}(z)
\,-\, \frac{\alpha_s(\mu_r)}{16\pi}(\rho^2+\bar{\rho}^2) \, E^2\, \log \frac{E^2}{\mu_r^2}
\right).\nonumber\\
 \label{eq:op_th3}
\end{eqnarray}

\noindent To further simplify the integrand we would like to select a natural value for the renormalisation scale that removes the $ (\rho \mu_r)^{b_0} (\bar{\rho} \mu_r)^{b_0}$ factor in the pre-exponent.
 Hence we choose
the value of $\mu_r$ to be set by the geometric average of the instanton sizes,
\begin{equation}
\mu_r \,=\, 1/\sqrt{\rho \bar{\rho}}\,,
\label{eq:mdef}
 \end{equation} 
 and as the result, all the running coupling constants appearing on the right hand side of \eqref{eq:op_th3} are given by the following 1-loop expression,
\begin{equation}
\frac{4\pi}{\alpha_s(1/\sqrt{\rho \bar{\rho}})} \,=\, \frac{4\pi}{\alpha_s(E)} \,-\, b_0 \log\left( \rho\bar{\rho} \,E^2\right)
\,.
\label{eq:alphadef}
 \end{equation} 

\bigskip
%%%%%%%%%%%%%%%%%%%%%%%%%%%%%%%%
\subsection{More on instanton--anti-instanton interaction}
%%%%%%%%%%%%%%%%%%%%%%%%%%%%%%%%
\medskip

It can be useful to separate the instanton--anti-instanton interaction potential $U_{\rm int}$ from the total action $S_{I\bar{I}}$,
\begin{equation}
U_{\rm int} (z) \,=\,  S_I\,+\, S_{\bar{I}}\,-\, S_{I\bar{I}}(z) \,=\, \frac{4\pi}{\alpha_s(\mu_r)} \, \left( 1-{\cal S}(z)\right)
\,,
\label{eq:Uintdef}
\end{equation}
where 
\begin{equation}
S_I\,=\, \frac{2\pi}{\alpha_s(\mu_r)} \,= \, S_{\bar{I}}\,,
\end{equation}
denote the individual actions of the single-instanton and the single-anti-instanton. It then follows from our earlier discussion that
in the limit of large separations, the interaction potential vanishes, and in the opposite limit where the individual instantons mutually annihilate,
the interaction cancels the effect of the individual instanton actions,
\begin{eqnarray}
\label{eq:Uintinfz}
\lim_{z\to \infty} U_{\rm int} &=& \frac{6}{z^2} +{\cal O}\left(\frac{1}{z^4}\log z  \right) \,\to\, 0\,,
\\
 \lim_{z\to 1} U_{\rm int} &=& 2 S_I\left( 1 - \frac{6}{5} (z-1)^2 +{\cal O}\left((z-1)^3 \right) \right)  \,\to\,  2 S_I\,.
\end{eqnarray}
The exponent of the instanton--anti-instanton action appearing in the optical theorem expression for the instanton total cross-section \eqref{eq:op_th3},
can be interpreted as the series expansion in powers of the instanton interaction potential,
\begin{equation}
\exp\left(-\frac{4\pi}{\alpha_s(\mu_r)} \,{\cal S}(z) \right)\,=\, 
\sum_{n=0}^{\infty}\frac{1}{n!} \, \left(U_{\rm int}\right)^n\, \exp\left(- S_{I} \,-\, S_{\bar{I}} \right)\,,
\label{eq:nsum}
\end{equation}
where $n$ is the number of the cut propagators in the imaginary part of the forward elastic scattering amplitude, i.e. the number of final state gluons
in the instanton process. The expression \eqref{eq:nsum} will be useful for in the following section for obtaining the mean number of final state gluons from our optical-theorem-based approach.

\medskip

We should further note that  the expression \eqref{eq:Szdef} given above corresponds to the action of the instanton--anti-instanton configuration for the choice of the relative orientation matrix $\Omega$ that corresponds to the maximal attraction between the instanton and the anti-instanton. In general one should integrate over all relative orientations on the right hand side of \eqref{eq:op_th3}. 
The result of this integration (see Appendix B) is,
\begin{eqnarray}
 \int d\Omega  \,e^{-\,\frac{4\pi}{\alpha_s(\mu_r)} \,{\cal S}(z, \Omega)} &=&
\frac{1}{9\sqrt{\pi}} \left(\frac{3}{U_{\rm int}}\right)^{7/2} \,e^{-\,\frac{4\pi}{\alpha_s(\mu_r)} \,{\cal S}(z)} \nonumber\\
 &=&
\frac{1}{9\sqrt{\pi}} \left(\frac{3\,\alpha_s(\mu_r)  }{4\pi(1-{\cal S}(z))}\right)^{7/2} \,e^{-\,\frac{4\pi}{\alpha_s(\mu_r)} \,{\cal S}(z)}
 \label{eq:op_Om}
\end{eqnarray}

\bigskip
%%%%%%%%%%%%%%%%%%%%%%%%%%%%%%%%%%%%%%%%%%%%%%%%%%%%
\subsection{The master integral}
\label{sec:master_int}
%%%%%%%%%%%%%%%%%%%%%%%%%%%%%%%%%%%%%%%%%%%%%%%%%%%%

We now introduce dimensionless integration variables,
\begin{eqnarray}
r_0 &=& R_0E\,, \qquad r \,=\, |\vec{R}|E\,,
\label{eq:resc1} \\
y &=& \rho\bar{\rho}\, E^2\,, \,\, \,\, \qquad x \,=\, \rho/\bar{\rho}\,,
\label{eq:resc2}
\end{eqnarray}
and use them to write down the instanton parton-level cross-section $\hat\sigma_{\rm tot}^{\rm inst}$ integral in \eqref{eq:op_th3}
in the form,
\begin{eqnarray}
\hat\sigma_{\rm tot}^{\rm inst} \,(E)  \,=\,\frac{1}{E^2}\, 
{\rm Im}\,  \int_{-\infty}^{+\infty} dr_0 \, e^{r_0}\, G(r_0, E)\,,
\label{eq:op_ssig1}
\end{eqnarray}
where
\begin{eqnarray}
G(r_0, E) \,=\,
 \frac{\kappa^2 \pi^4 }{2^{17}}\sqrt{\frac{\pi}{3}} \, \int_0^{\infty} r^2\, dr  &&\int_0^{\infty} \frac{d x}{x} \int_0^{\infty} \frac{d y}{y}
 \left( \frac{4\pi}{\alpha_s}\right)^{21/2} \left(\frac{1}{1-{\cal S}(z)}\right)^{7/2}\,
% \times 
\nonumber\\ \nonumber
  \\
 {\cal K}_{\rm ferm}(z)
 &&\exp\left(-\,\frac{4\pi}{\alpha_s} \,{\cal S}(z)
\,-\, \frac{\alpha_s}{4\pi}\,  \frac{x+1/x}{4}\,y\log y 
\right).
 \label{eq:op_ssig2}
\end{eqnarray}

\noindent Here $\kappa$, ${\cal S}(z)$ and $ {\cal K}_{\rm ferm}(z)$ are given by \eqref{eq:kapdef}, \eqref{eq:Szdef} and \eqref{eq:Kferm}-\eqref{eq:omegaF},
and the conformal ratio variable $z$ is expressed in terms of our dimensionless variables via,
\begin{equation}
z\,=\, \frac{1}{2} (\xi +(\xi^2-4)^{1/2})\,, \quad {\rm where} \quad
\xi\,=\, \frac{r_0^2+r^2}{y} + x +\frac{1}{x}\,,
\end{equation}
in agreement with the the expression \eqref{eq:zdef}.

\medskip
The final ingredient we need is the expression \eqref{eq:alphadef} for the running couplings  in terms of the $y$ variable, 
\begin{eqnarray}
\frac{4\pi}{\alpha_s} (y;E)&=& \frac{4\pi}{\alpha_s(E)} \,-\, b_0 \log y
\nonumber\\
&=& \frac{4\pi}{0.416} \,+\, 2b_0 \log \frac{E}{1{\rm GeV}} \,-\, b_0 \log y
\,,
\label{eq:alphay}
 \end{eqnarray} 
as follows from \eqref{eq:alphadef} and \eqref{eq:resc2}. We will thus set $\frac{4\pi}{\alpha_s} = \frac{4\pi}{\alpha_s} (y;E)$ in the integrand \eqref{eq:op_ssig2} (including the function in the exponent and the non-exponential terms in the integrant in \eqref{eq:op_ssig2} ).

\medskip

To compute the instanton cross-section  \eqref{eq:op_ssig1}  
we first numerically evaluate the integral \eqref{eq:op_ssig2} and obtain the values for $G(r_0, E) $
for a wide range of both arguments, $r_0$ and $E$.
After that we 
perform the final integration over $r_0$ in \eqref{eq:op_ssig1} by expanding the integrand in
\begin{eqnarray}
\hat\sigma_{\rm tot}^{\rm inst} \,(E)  \,=\,\frac{1}{E^2}\, 
{\rm Im}\,  \int_{-\infty}^{+\infty} dr_0 \, e^{r_0 + \log G(r_0, E)}\,,
\label{eq:op_ssig1_fin}
\end{eqnarray}
around the 
stationary point solution for $r_0$ of the function $r_0 + \log G(r_0, E) $ in the exponent,
\begin{equation}
r_0(E): \quad \partial_{r_0} \log G(r_0, E) \,=\, -1\,,
\label{eq:r0sp}
\end{equation}
for each value of $E$.
The saddle-point evaluation of the $r_0$ integral \eqref{eq:op_ssig1_fin} gives,
\begin{eqnarray}
\hat\sigma_{\rm tot}^{\rm inst} \,(E) &\approx& \frac{1}{E^2}\, \sqrt{\frac{2\pi}{-\partial_{r_0}^2\log G}}\biggr\vert_{r_0=r_0(E)}\,
 e^{r_0(E) + \log G(r_0(E), E)} \nonumber\\
  &=& \frac{1}{E^2}\, \sqrt{\frac{2\pi}{W''}}\biggr\vert_{r_0=r_0(E)}\,
 e^{r_0(E) -W(r_0(E), E)} \,,
\label{eq:op_ssig1_fin2}
\end{eqnarray}
where we have defined,
\begin{equation}
W(r_0, E) \,= \, - \log G_0(r_0,E) \,, \qquad W'(r_0, E) \,=\, - \partial_{r_0} \log G(r_0, E)\,.
\end{equation}

\begin{figure}
\includegraphics[width=0.5\textwidth]{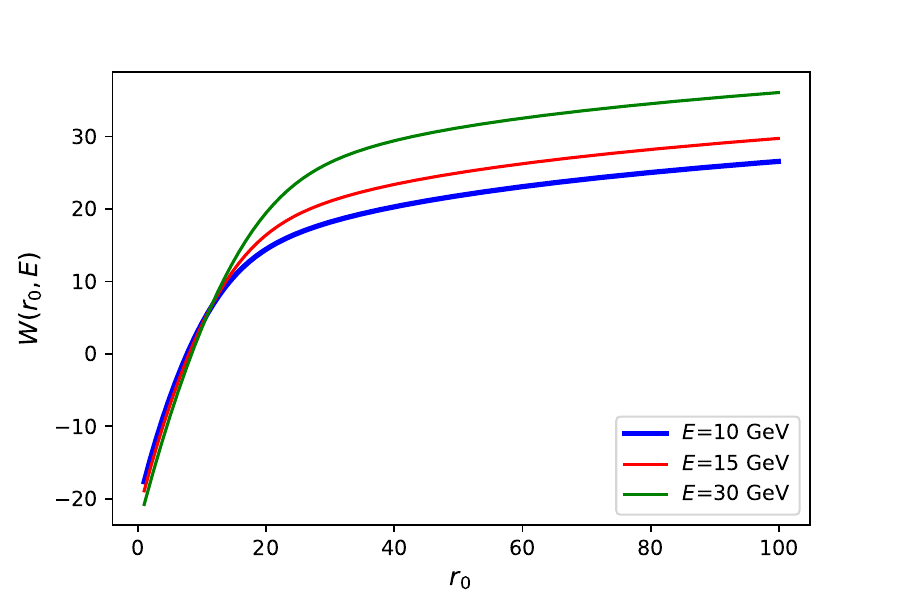}
\includegraphics[width=0.5\textwidth]{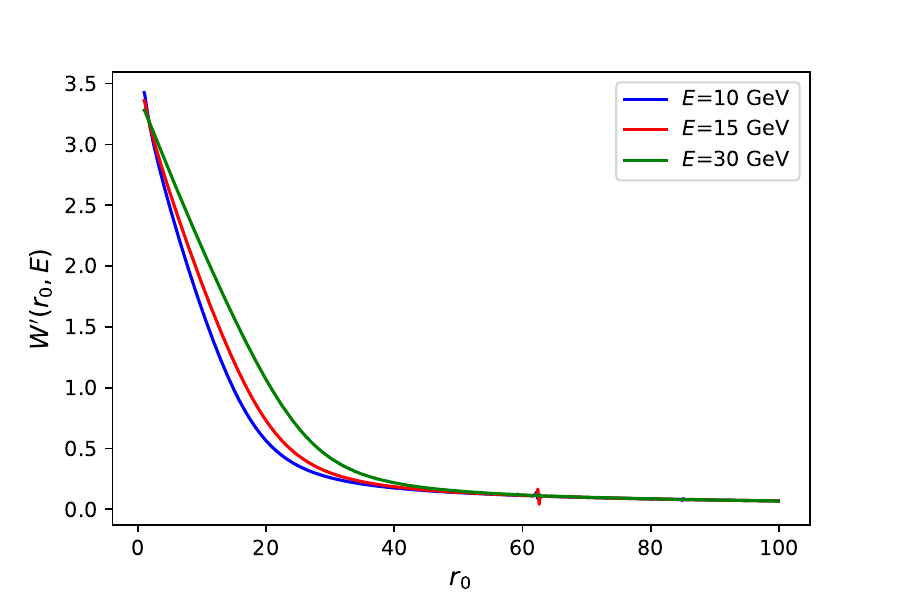}
\caption{(left) $W(r_0, E)$ plotted for $E$=10, 15, 30 GeV and 0<$r_{0}$<100. (right) $W'(r_0, E)$ plotted for $E$=10, 15, 30 GeV and 0<$r_{0}$<100. NB one should ignore the small spike at $r_{0}$=60 as this is merely an artefact of the numerical accuracy of our differentiation and integration functions.}
\label{fig:Wplot}
\end{figure}

The numerical integration in \eqref{eq:op_ssig2} was carried out using the python package SciPy\cite{2020SciPy-NMeth}
for $E$ in the range in $10 < E < 2000 $ GeV and for a wide range in $r_0$ to accommodate a sufficiently large interval
around the expected values of the saddle-point $r_0(E)$ in \eqref{eq:r0sp}.
In Fig.\ref{fig:Wplot} we plot the resulting functions  
$W(r_0, E)$ and $W'(r_0, E)$
for fixed values of $E= 10, 15, 30$~GeV in the range $0<r_0 <100$ alongside the $r_0$.
The function $W(r_0, E)$ plays the role of the  effective instanton-anti-instanton Euclidean action (this is because it arises from integrating
the exponent of the classical action
$e^{-\frac{4\pi}{\alpha_s} \,{\cal S}(z)}$ over the collective coordinates of non-negative modes of the $I\bar{I}$ configuration 
on the r.h.s. of \eqref{eq:op_ssig2}).
The saddle-point value for $r_0$ is given by the equation $W'(r_0, E) =1$ for each fixed value of $E$,
as dictated by \eqref{eq:r0sp} above.

Having determined $W(r_0, E)$ and its derivatives as functions of $r_0$ and $E$ we can now carry out the final integration over $r_0$ 
using the saddle-point approximation formula \eqref{eq:op_ssig1_fin2} for the imaginary part of the forward elastic scattering amplitude and hence for the 
partonic instanton cross-section $\hat\sigma_{\rm tot}^{\rm inst} \,(E)$.
Our final results for the partonic instanton cross-section \eqref{eq:op_ssig1}
are displayed in Table~\ref{Table:MainResult}.

Then the hadronic cross-sections are calculated from these partonic cross-sections using the NNPDF3.1luxQED NNLO dataset with $\alpha_{s}\left(M_{Z}\right)=0.118$ \cite{Buckley:2014ana} \cite{Bertone:2017bme} and displayed in Table \ref{Table:HadXSection}.  These are calculated using the usual formula
\begin{equation}
\sigma_{pp\rightarrow I}\left(\hat{s}>\hat{s}_{\rm min}\right)=\int_{\hat{s}_{min}}^{s_{pp}} dx_{1}dx_{2}\quad f\left(x_{1},Q^{2}\right)f\left(x_{2},Q^{2}\right)\hat{\sigma}\left(\hat{s}=x_{1}x_{2}s_{pp}\right)
\label{eq:Ihadroinic}
\end{equation}
where $s_{pp}$ is the centre-of-mass energy of the hadron collider, $\hat{\sigma}$ is the partonic instanton cross-section and $\hat{s}_{\rm min}$ is the minimum invariant mass squared of the produced system. NB here we are only considering the gluon initiated process, otherwise we require a sum over such integrals.

\begin{table}[]
\centering
\scalebox{0.9}{
\begin{tabular}{|l|l|l|l|l|l|l|l|l|}
\hline
$\sqrt{\hat{s}}$ [GeV] &   50                               & 100                            & 150   & 200   & 300              & 400             & 500                   \\ 
\hline
$\langle n_{g}\rangle$ & 9.43                             & 11.2                           & 12.22 & 12.94 & 13.96            & 14.68           & 15.23          \\
\hline
$\hat\sigma_{\rm tot}^{\rm inst}$ [pb] & 207.33$\times10^{3}$ & 1.29$\times10^{3}$ & 53.1  & 5.21  & 165.73$\times10^{-3}$ & 13.65$\times10^{-3}$ & 1.89$\times10^{-3}$    \\
\hline
\end{tabular}
}
\caption{The instanton cross-section presented for a range of partonic C.o.M. energies $\sqrt{\hat{s}}=E$ and the mean number of gluons at this energy calculated using 
Eq.~\eqref{eq:numbergluons}.}
\label{Table:MainResult}
\end{table}

\begin{table}[]
\centering
\scalebox{0.9}{
\begin{tabular}{|l|l|l|l|l|l|l|l|l|l|l|}
\hline
$E_{\rm min}$ [\GeV]                              & 50          & 100        & 150        & 200        & 300          & 400       & 500 \\ 
\hline
$\sigma_{p\bar{p}\rightarrow I}$ &  2.62 $\mu$b & 2.61 nb & 29.6 pb & 1.59 pb & 6.94 fb & 105 ab & 3.06 ab  \\
$\sqrt{s_{p\bar{p}}}$=1.96 TeV &&&&&&&\\
\hline
$\sigma_{pp\rightarrow I}$ &  58.19 $\mu$b & 129.70 nb & 2.769 nb & 270.61 pb & 3.04 pb & 114.04 fb & 8.293 fb \\
$\sqrt{s_{pp}}$=14 TeV &&&&&&&\\
\hline
$\sigma_{pp\rightarrow I}$ &  211.0 $\mu$b & 400.9 nb & 9.51 nb & 1.02 nb & 13.3 pb & 559.3 fb & 46.3 fb  \\
$\sqrt{s_{pp}}$=30 TeV &&&&&&&\\
\hline
$\sigma_{pp\rightarrow I}$ &  771.0 $\mu$b & 2.12 $\mu$b & 48.3 nb & 5.65 nb & 88.3 pb & 4.42 pb & 395.0 fb  \\
$\sqrt{s_{pp}}$=100 TeV &&&&&&&\\
\hline
\end{tabular}
}
\caption{Hadronic cross-sections for QCD instanton processes at a range of colliders with different C.o.M. energies $\sqrt{s_{p\bar{p}}}$ evaluated using  
Eq.~\eqref{eq:Ihadroinic}. The minimal allowed partonic energy is $E_{\rm min} = \sqrt{\hat{s}_{\rm min}}$. }
\label{Table:HadXSection}
\end{table}

\bigskip
%%%%%%%%%%%%%%%%%%%%%%%%%%%%
\subsection{Mean number of final state gluons}
\label{sec:n_gluons}
%%%%%%%%%%%%%%%%%%%%%%%%%%%%
\medskip

In our approach of computing the total partonic cross-section via the optical theorem in \eqref{eq:op_ssig1}, \eqref{eq:op_ssig2} we have already effectively
summed over the number gluons $n_g$ in the final state. This sum can be uncovered by using the series expansion \eqref{eq:nsum}
 of the
exponent of the instanton--anti-instanton action on the right hand side of \eqref{eq:op_ssig2},
\begin{eqnarray}
G(r_0, E) \,=\,
 \frac{\kappa^2 \pi^4 }{2^{17}}\sqrt{\frac{\pi}{3}} && \int_0^{\infty} r^2\, dr \int_0^{\infty} \frac{d x}{x} \int_0^{\infty} \frac{d y}{y}
 \left( \frac{4\pi}{\alpha_s}\right)^{21/2} \left(\frac{1}{1-{\cal S}(z)}\right)^{7/2}\,
  {\cal K}_{\rm ferm}(z)
% \times 
\nonumber\\ \nonumber
  \\
&&\,
\sum_{n_g=0}^{\infty}\frac{1}{n_g!} \, \left(U_{\rm int}\right)^{n_g}\, \exp\left(-\,\frac{4\pi}{\alpha_s} 
\,-\, \frac{\alpha_s}{4\pi}\,  \frac{x+1/x}{4}\,y\log y 
\right).
 \label{eq:op_ssig2sum}
\end{eqnarray}
The mean value of $n_g$ (i.e. the value that gives the dominant contribution to the integral) is then easily found to be given by the expectation value of the 
interaction potential,
\begin{equation} 
\langle n_g\rangle  \,=\, \langle U_{\rm int} \rangle
\,,
\end{equation}
where the expectation value of $ \langle U_{\rm int} \rangle$ is obtained by inserting 
$U_{\rm int} = \frac{4\pi}{\alpha_s(y;E)} \, \left( 1-{\cal S}(z)\right)$ into the integrand on the right hand side of \eqref{eq:op_ssig1}, \eqref{eq:op_ssig2} and normalising by 
$1/(E^2 \hat\sigma_{\rm tot}^{\rm inst} )$.

In practice, we compute
\begin{eqnarray}
\label{eq:numbergluons}
\langle n_g \rangle &=& \frac{1}{G(r_0, E)} \,
 \frac{\kappa^2 \pi^4 }{2^{17}}\sqrt{\frac{\pi}{3}} \, \int_0^{\infty} r^2\, dr  \int_0^{\infty} \frac{d x}{x} \int_0^{\infty} \frac{d y}{y}
 \left( \frac{4\pi}{\alpha_s(y;E)}\right)^{21/2} \left(\frac{1}{1-{\cal S}(z)}\right)^{7/2}\,
% \times 
\\ \nonumber
  \\
&& 
  {\cal K}_{\rm ferm}(z)\frac{4\pi}{\alpha_s(y;E)} \, \left( 1-{\cal S}(z)\right) \,
 \cdot \exp\left(-\,\frac{4\pi}{\alpha_s(y;E)} \,{\cal S}(z)
\,-\, \frac{\alpha_s(y;E)}{4\pi}\,  \frac{x+1/x}{4}\,y\log y 
\right)\,.\nonumber
\label{eq:vev_ng_U}
\end{eqnarray}
On the right hand side we have integrated over the $y,x,r$ variables.
The variable $r_0$ is taken to be at its saddle-point value  
for each fixed value of the energy $E$.

To account for the possibility of the new shifted saddle-point we do this:
\begin{eqnarray}
\langle n_g \rangle &=& \frac{1}{{\rm Im}\,  \int_{-\infty}^{+\infty} dr_0 \, e^{r_0}\, G(r_0, E)} \times\, 
\label{eq:vev_ng_U2}
\\
&& \quad {\rm Im}\,  \int_{-\infty}^{+\infty} dr_0 \, e^{r_0}\,
 \frac{\kappa^2 \pi^4 }{2^{17}}\sqrt{\frac{\pi}{3}} \, \int_0^{\infty} r^2\, dr  \int_0^{\infty} \frac{d x}{x} \int_0^{\infty} \frac{d y}{y}
 \left( \frac{4\pi}{\alpha_s(y;E)}\right)^{21/2} \left(\frac{1}{1-{\cal S}(z)}\right)^{7/2}\,
% \times 
\nonumber\\ \nonumber
  \\
&& \quad
 {\cal K}_{\rm ferm}(z) \frac{4\pi}{\alpha_s(y;E)} \, \left( 1-{\cal S}(z)\right) \,
 \cdot \exp\left(-\,\frac{4\pi}{\alpha_s(y;E)} \,{\cal S}(z)
\,-\, \frac{\alpha_s(y;E)}{4\pi}\,  \frac{x+1/x}{4}\,y\log y 
\right).
\nonumber
\end{eqnarray}

\medskip

%%%%%%%%%%%%%%%%%%%%%%%%%%%%%%%%%%%%%%%%%%%%%%%%%%%%
\medskip
\section{\label{Sec:Recoil}Instanton Recoil by a Jet}

In this section we explain how to generalise the calculation of the instanton process presented above to the case where a jet
is emitted from one of the initial state partons.  This is of course an important process for collider studies as it allows one to recoil the instanton-generated multi-particle final state by a high-$p_T$ jet.

When the jet is carrying momentum $p$ produced from an initial parton $p_1$, the secondary gluon $q$ entering the instanton vertex will necessarily have a virtuality $q^2=-Q^2\neq 0$.\footnote{In the complimentary scenario where a high-$p_T$ jet is emitted from the instanton vertex in the
{\it final state}, no virtualities arise, all momenta entering and leaving the instanton vertex are on-shell, and the formalism presented in the earlier section requires no modicfications.}
In the partonic centre of mass frame we have,
\begin{eqnarray}
p_1 &=& (\sqrt{\hat{s}}/2, 0, p_L)\,, \quad p_2 \,=\, (\sqrt{\hat{s}}/2, 0, -p_L)\,, \quad {\rm where\,\,} |p_L|=\sqrt{\hat{s}}/2\,, \nonumber\\
p_1 &=& q + p \,, \quad p\,=\, (|p_T|, p_T, 0) \,, \quad Q^2 =-q^2 = -(p_1-p)^2=\, \sqrt{\hat{s}} \, p_T\,.
\end{eqnarray}
Here we have assumed for simplicity that the jet momentum $p$ is transverse, i.e. it does not have a longitudinal component.

The kinematic-invariant CoM energy for the parton-level process is, as before, $\sqrt{\hat{s}}$, where $\hat{s}=(p_1+p_2)^2$. On the other hand,
the invariant mass entering the instanton vertex $\sqrt{s'}$ is now different,
\begin{equation}
s' \,=\, (q+p_2)^2 \,=\, \hat{s}-2Q^2 \,=\, \sqrt{\hat{s}}\,(\sqrt{\hat{s}}-2p_T)\,.
\label{eqn:snewhat}
\end{equation}

The virtuality $Q$ of an incoming gluon leg, induced by a no-zero $p_T$, introduces a multiplicative form-factor $e^{-\,Q\rho}$ into the instanton vertex.
This is a well-known result \cite{Khoze:1990bm,Balitsky:1992vs,Ringwald:1998ek} that is a direct consequence of Fourier transforming the instanton field 
to the momentum space to obtain $A^{\rm inst}_{LSZ}(q)$, where the momentum $q$ has a large virtuality, $Q^2$.
For the instanton cross-section one needs to compute,
$A^{\rm inst}_{LSZ}(q)\,A^{\rm inst}_{LSZ}(p_2)\,A^{\overline{\rm inst}}_{LSZ}(-q)\,A^{\overline{\rm inst}}_{LSZ}(-p_2)$, in analogy with Eq.~\eqref{eq:LSZA},
which gives the overall form-factor,
\begin{equation}
\exp\left(-Q (\rho+\bar{\rho})\right)\,=\, \exp\left(-\frac{Q}{E} \sqrt{y\left(x+1/x+2\right)}\right),
\end{equation}
that needs to be included in the integral \eqref{eq:op_th3}. On the right hand side of this equation we used our standard dimensionless variables $x$ and $y$
defined in~\eqref{eq:resc1}-\eqref{eq:resc2}.

The second modification of the integral in \eqref{eq:op_th3}, is that the the energy variable $E$ corresponds to the instanton vertex energy $E=\sqrt{s'}$ defined in \eqref{eqn:snewhat}, which is smaller than the overall invariant mass $\sqrt{\hat{s}}$ of the parton-level process.

\medskip

In summary, the instanton parton-level cross-section $\hat\sigma_{\rm tot}^{\rm inst} \,(\sqrt{\hat{s}},p_T)$ is computed as follows:
\begin{enumerate}
\item For each pair of physical variables $\hat{s}$, $p_T$, introduce the auxiliary variables $E$ and $Q$,
\begin{equation}
Q^2 = p_T \sqrt{\hat{s}} \,, \qquad E^2=\hat{s}-2Q^2\,.
\end{equation}

\item Numerically compute the integral,
\begin{eqnarray}
\tilde{G}(r_0, E,Q) \,=\,
 \frac{\kappa^2 \pi^4 }{2^{17}}\sqrt{\frac{\pi}{3}} \, \int_0^{\infty} r^2\, dr  \int_0^{\infty} \frac{d x}{x} \int_0^{\infty} \frac{d y}{y}
 \left( \frac{4\pi}{\alpha_s}\right)^{21/2} \left(\frac{1}{1-{\cal S}(z)}\right)^{7/2}\,
% \times 
\nonumber\\ \nonumber
  \\
 {\cal K}_{\rm ferm}(z)
 \exp\left(-\,\frac{4\pi}{\alpha_s} \,{\cal S}(z)
\,-\, \frac{\alpha_s}{4\pi}\,  \frac{x+1/x}{4}\,y\log y 
\,-\, \frac{Q}{E} \sqrt{y\left(x+\frac{1}{x}+2\right)}
\right)
 \label{eq:op_ssig2_Q}
\end{eqnarray}
and use it to evaluate the expression for the cross-section,
\begin{eqnarray}
I (E,Q)  \,=\,\frac{1}{E^2}\, 
{\rm Im}\,  \int_{-\infty}^{+\infty} dr_0 \, e^{r_0}\, \tilde{G}(r_0, E,Q)\,,
\label{eq:op_ssig1_Q}
\end{eqnarray}
in the saddle-point approximation, as before. 
\item The cross-section in physical variables is then obtained via,
\begin{eqnarray}
\hat\sigma_{\rm tot}^{\rm inst} \,(\sqrt{\hat{s}},p_T) \,=\,
I (E,Q)\Bigl\vert_{Q^2 = p_T \sqrt{\hat{s}} \,,\, E^2=\hat{s}-2p_T \sqrt{\hat{s}}}
\label{eq:op_ssig1_final}
\end{eqnarray}
\end{enumerate}

Table~\ref{Table:MainResultQ} presents the results for the instanton cross-section at parton level for a range of partonic CoM energies
$\sqrt{\hat{s}}$ and for a fixed value of the recoiled jet transverse momentum $p_T=150$ GeV. The resulting cross-sections fixed are negligibly small.
To complement these results we have also computed instanton cross-sections for the case where $p_{T}$ is scaled with the energy.
Table~\ref{Table:VariablepT} presents the results at parton level where the recoiled jet transverse momentum is chosen as $p_{T}={\sqrt{\hat{s}}}/{3}$.

\begin{table}[]
\centering
\scalebox{0.9}{
\begin{tabular}{|l|l|l|l|l|l|l|l|l|}
\hline
$\sqrt{\hat{s}}$ [GeV]&310 &   350   &375                            & 400                            & 450   & 500                  \\ 
\hline
$\hat\sigma_{\rm tot}^{\rm inst}$  [pb] & 3.42$\times10^{-23}$ & 1.35$\times10^{-18}$&1.06$\times10^{-17}$ & 1.13$\times10^{-16}$ & 9.23$\times10^{-16}$  & 3.10$\times10^{-15}$  \\
\hline
\end{tabular}
}
\caption{The instanton partonic cross-section recoiled against a hard jet with $p_T=150$ GeV emitted from an initial state and calculated using
Eq.~\eqref{eq:op_ssig1_final}.
Results for the cross-section are shown  for a range of partonic C.o.M. energies $\sqrt{\hat{s}}$.}
\label{Table:MainResultQ}
\end{table}

\begin{table}[]
\centering

\scalebox{0.9}{
\begin{tabular}{|l|l|l|l|l|l|l|l|l|}
\hline
$\sqrt{\hat{s}}$ [GeV]                           & 100                            & 150   & 200   & 300              & 400             & 500                   \\ 
\hline
$\hat\sigma_{\rm tot}^{\rm inst}$ [pb] & 1.68$\times10^{-7}$ & 1.20$\times10^{-9}$  & 3.24$\times10^{-11}$  & 1.84$\times10^{-13}$ & 4.38$\times10^{-15}$ & 2.38$\times10^{-16}$    \\
\hline
\end{tabular}
}
\caption{The cross-section presented for a range of partonic C.o.M. energies $\sqrt{\hat{s}}=E$ where the recoiled $p_{T}$ is scaled with the energy,
$p_{T}=\sqrt{\hat{s}}/{3}$.}
\label{Table:VariablepT}
\end{table}

From the results in Tables~\ref{Table:VariablepT} and \ref{Table:MainResultQ} we see that the cross-sections calculated for the processes where the instanton recoils against a jet with large momentum are too small to be observable at any present or envisioned high-energy collider. While increasing the transverse momentum for objects that are difficult to reconstruct by recoiling them against a hard object is often a popular method to improve the sensitivity of the LHC to new physics, see e.g. \cite{Kribs:2009yh,Schlaffer:2014osa,Harris:2014hga,Harris:2015kda}, the instanton shields itself from such an an option. Consequently, the only way to obtain sensitivity to instantons is to disentangle their spherical radiation profile, made of fairly soft jets, from SM QCD backgrounds.

\bigskip
%%%%%%%%%%%%%%%%%%%%%%%%%%%%%%%%%
\section{Search for Instanton Events at Hadron Colliders}
\label{sec:analysis}
%%%%%%%%%%%%%%%%%%%%%%%%%%%%%%%%%

%%%%%%%%%%%%%%%%%%%%%%%%%%%%%%%%%
\subsection{Topology of Instanton Events}
\label{sec:topology}
%%%%%%%%%%%%%%%%%%%%%%%%%%%%%%%%%
\medskip

Since the global event topology of instanton processes is spherically symmetric, and therefore distinctly different from perturbative-QCD events, event shape observables \cite{Banfi:2004nk} can be a powerful way to identify these processes. 

The transverse sphericity tensor is defined as
\begin{equation}
S^{\alpha\beta}=\frac{\sum_{i} p_{i}^{\alpha}p_{i}^{\beta}}{\sum _{i} \left\lvert \bold{p}_{i}^{2}\right\rvert},
\end{equation}
where $\alpha,\beta$ run over spatial indices and $i$ runs over the number of particles. Here $p_{i}$ is the two-dimensional transverse component of momentum. The transverse sphericity observable is then defined as $S=\frac{2\lambda_{2}}{\lambda_{1}+\lambda_{2}}$ where $\lambda_{i}$ are the eigenvalues of the transverse sphericity tensor and $\lambda_{2}<\lambda_{1}$. Transverse sphericity takes values between 0 and 1 with higher values denoting a higher degree of spherical symmetry. Therefore we would expect instanton processes to have a higher transverse sphericity than background processes which in general have some angular dependence. 

Spherocity is defined as 
\begin{equation}
S_{0}=\frac{\pi^{2}}{4}\min_{\vec{n}}\left(\frac{\sum_{i} \left\lvert\vec{p}_{\perp,i}\times\vec{n}\right\rvert}{ \sum_{i}\left\lvert\vec{p}_{\perp,i}\right\rvert}\right)^{2},
\end{equation}
where $\vec{n}$ is a unit vector with zero longitudinal component. Again, $S_0$ takes values between 0 and 1, with 1 representing a completely isotropic event and 0 being a pencil-like event. This variable is closely related to thrust which is defined as 
\begin{equation} 
\tau=1-\max_{\vec{n}}\frac{\sum_{i} \left\lvert\vec{p}_{i} \cdot \vec{n}\right\rvert}{ \sum_{i}\left\lvert\vec{p}_{i}\right\rvert},
\end{equation} 
where $\vec{n}$ is a unit vector. Thrust is 0 for pencil-like events and 0.5 for spherically symmetric events. The vector $\vec{n}$ which maximises this expression is known as the thrust axis.

The final shape variable we consider is broadening. The thrust axis automatically divides the event into a left hemisphere, $\mathcal{L}$ and a right hemisphere, $\mathcal{R}$. Left and right broadening is then defined as
\begin{equation}
\mathcal{B}_{\mathcal{L}}=\sum_{i\in\mathcal{L}}\frac{\lvert\vec{p}_{i}\times\vec{n}|}{\sum_{i}\lvert\vec{p}_{i}\rvert} \quad \mathcal{B}_{\mathcal{R}}=\sum_{i\in\mathcal{R}}\frac{\lvert\vec{p}_{i}\times\vec{n}|}{\sum_{i}\lvert\vec{p}_{i}\rvert}
\end{equation}
where $\vec{n}$ is the thrust axis. Total broadening $\mathcal{B}$ is then the sum of the left and right broadening, $\mathcal{B} = \mathcal{B}_{\mathcal{L}} + \mathcal{B}_{\mathcal{R}}$, and takes values between 0 and 0.5 with 0.5 being spherically symmetric. 

To show the different shapes for these observables between various perturbative SM processes and instanton events at the LHC and the Tevatron, we generate the background events using Pythia~8 \cite{Sjostrand:2014zea}. For the perturbative SM processes we consider the ones with largest cross-section and jet-rich final states, i.e. high and low-pT multi-jet events, min-bias events, $t\bar{t}$ production and $W$+jets events.
%We decided to look at experimental analyses to see if any constraints could be placed on the instanton cross-section. Our pileup plots come from summing over n minimum bias events, where n is Poisson-distributed with a mean of 34(2) at the LHC(Tevatron). 
For the signal we use \Rambo \cite{Kleiss:1985gy} to populate the phase space of the instanton final state. Each event contains four $q\bar{q}$ pairs and a poisson-distributed number of gluons, with a mean in accordance to $n_{g}$ in Table \ref{Table:MainResult}. 

%We generated 100,000 events with their energies weighted by the hadronic cross-sections and a minimum energy of 100 \GeV, as lower energies had almost all particles removed by our requirements on jet $E_{T}$. 

All processes are analysed using Fastjet \cite{Cacciari:2011ma}.
For the LHC we reconstruct jets using the anti-$k_{T}$ algorithm \cite{Cacciari:2008gp} with a cone-size of $R=0.4$ and $p_{T} \geq 10$~GeV. 
%reflecting the lowest $p_{T}$ trigger we found \citep{ATLAS:2016qun}. 
At the Tevatron  jets were analysed using the $k_{T}$ algorithm \cite{Cacciari:2008gp} with a cone-size of $R=0.7$ and were required to have 
$p_{T} \geq 5$~GeV.
%reflecting the lowest $p_{T}$ trigger we found and also the algorithms more commonly used in the Tevatron analyses . 
Leptons are required to have $p_{T} \geq 0.5$ GeV. It should be noted that the instanton processes are not showered or hadronised, but this should not significantly affect the analysis as the position and energy of the reconstructed jets are conserved to a good accuracy.

\begin{figure}[h]
\includegraphics[width=0.5\textwidth]{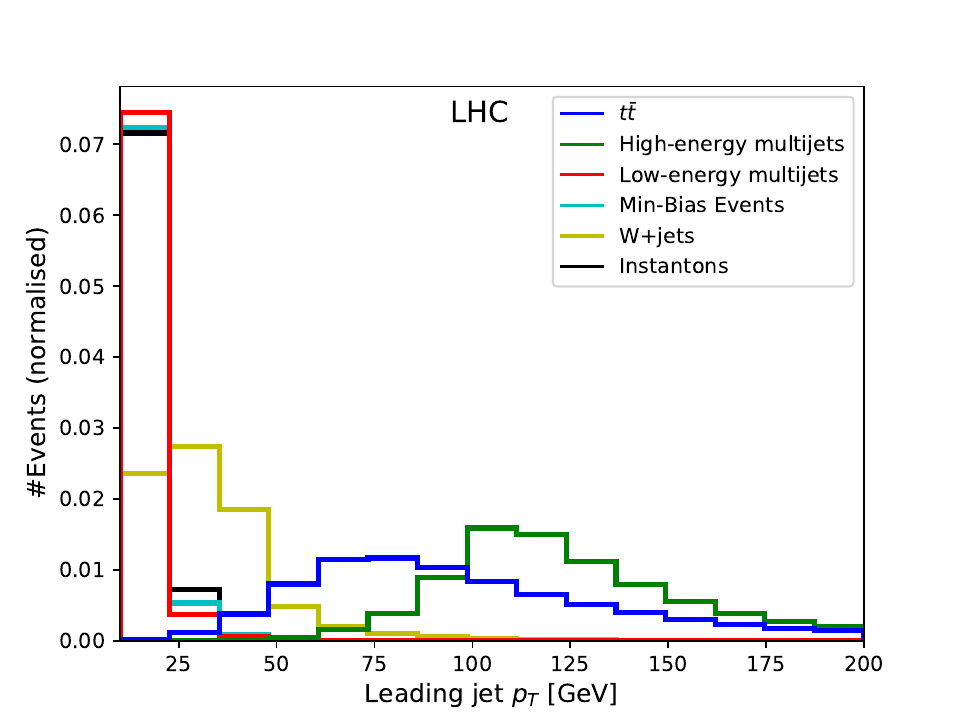}
\includegraphics[width=0.5\textwidth]{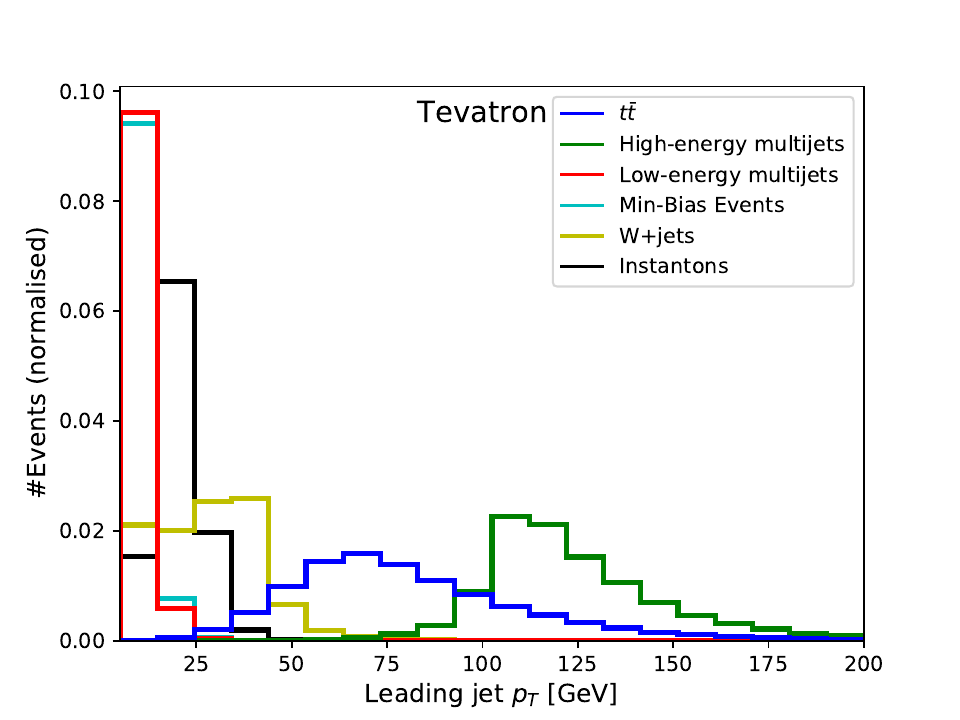}
\caption{The distribution of the $p_{T}$ of the leading jet for our background processes and instantons at the LHC (left) and  Tevatron (right).}
\label{fig:leadpt}
\end{figure}

\begin{figure}[h]
\includegraphics[width=0.5\textwidth]{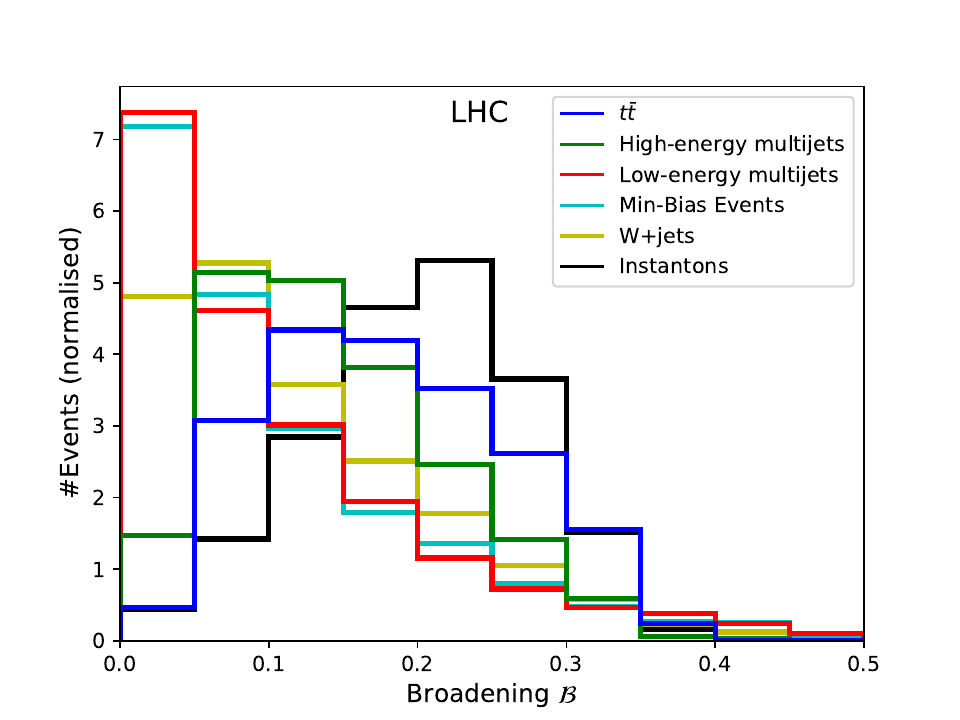}
\includegraphics[width=0.5\textwidth]{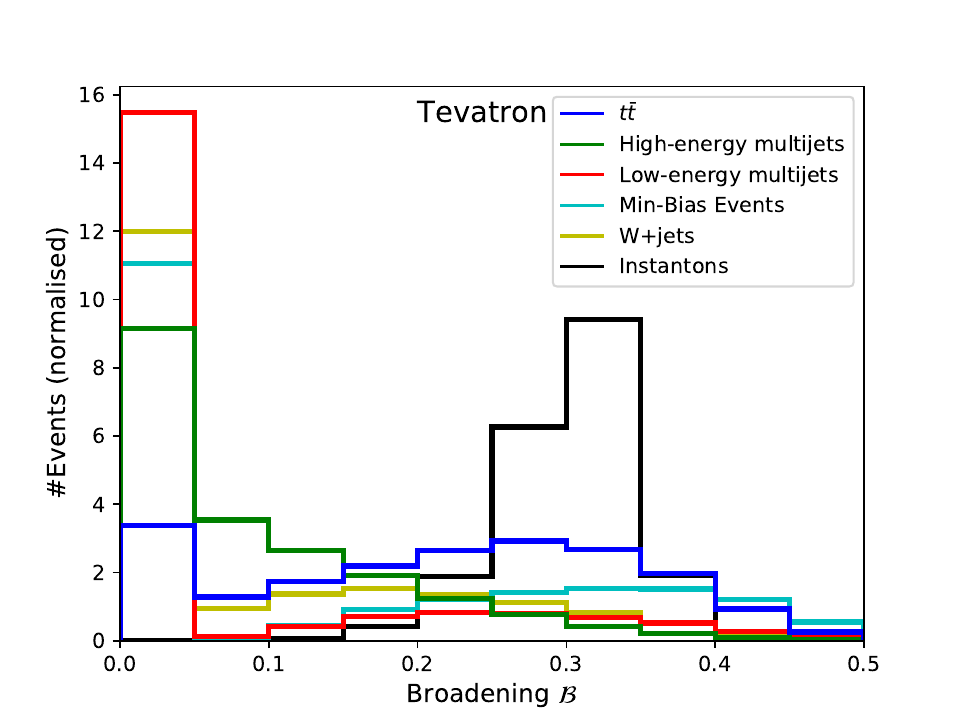}
\caption{The distribution of the broadening of events for our background processes and instantons at the LHC (left) and Tevatron (right).}
\label{fig:broadening}

\end{figure}

\begin{figure}[h]
\includegraphics[width=0.5\textwidth]{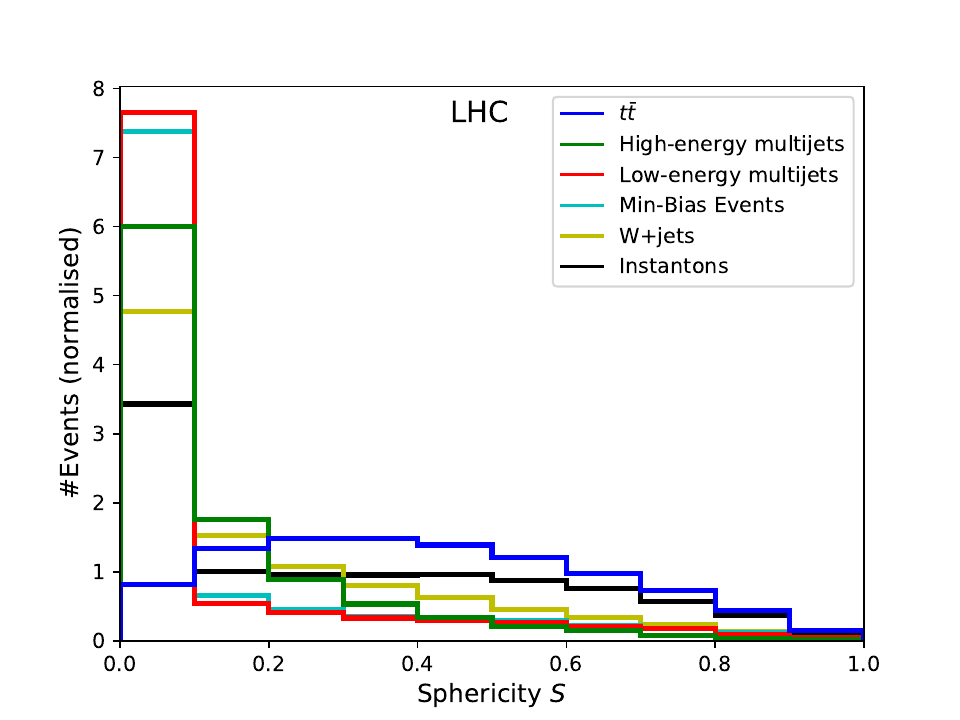}
\includegraphics[width=0.5\textwidth]{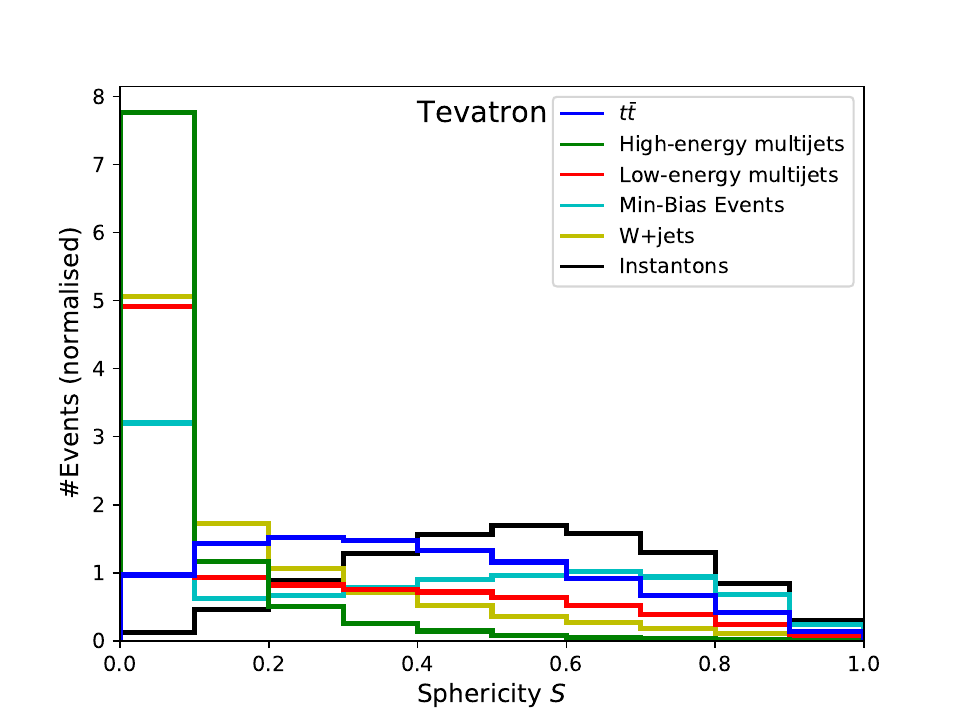}
\caption{The distribution of the transverse sphericity of events for our background processes and instantons at the LHC (left) and  Tevatron (right).}
\label{fig:sphericity}
\end{figure}

\begin{figure}[h]
\includegraphics[width=0.5\textwidth]{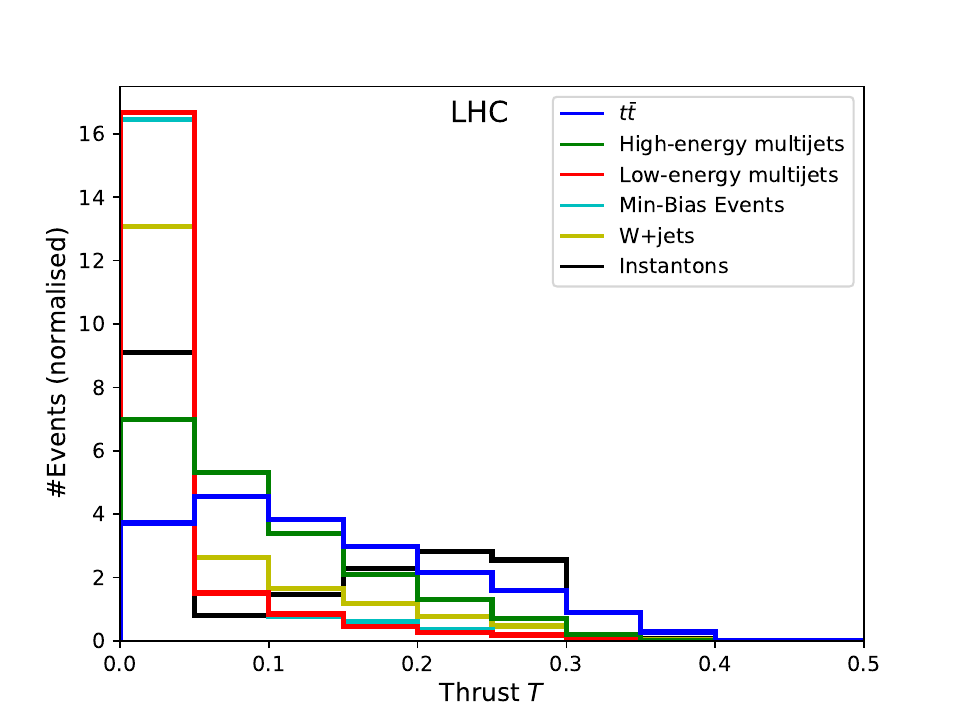}
\includegraphics[width=0.5\textwidth]{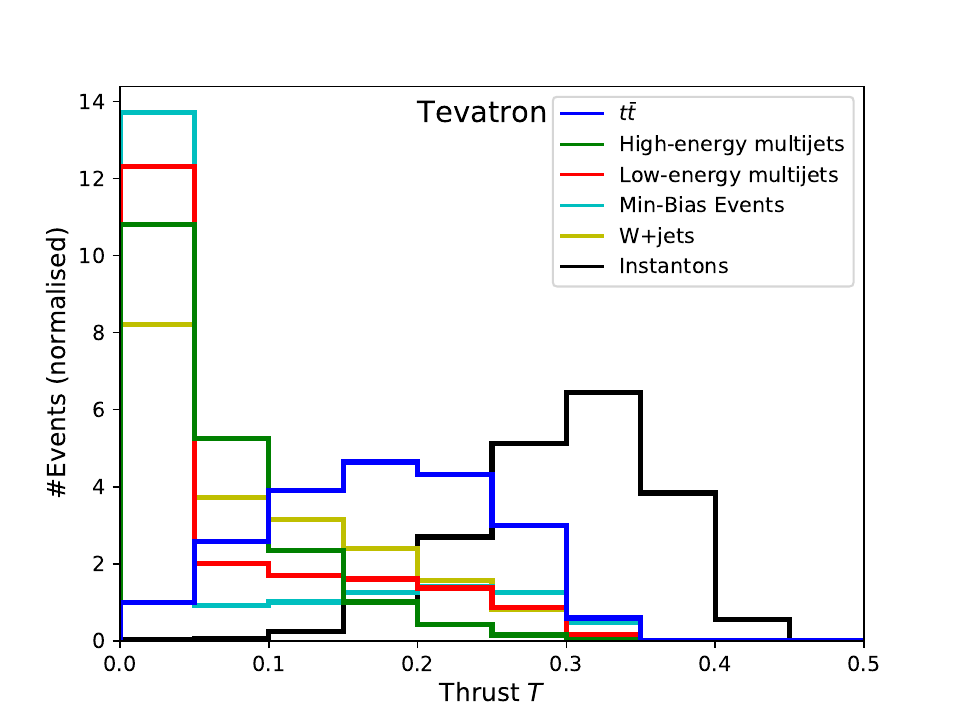}
\caption{The distribution of the thrust of events for our background processes and instantons at the LHC (left) and  Tevatron (right).}
\label{fig:thrust}
\end{figure}

We show in Fig.~\ref{fig:leadpt} the distribution for the $p_T$ of the leading jet, in Fig.~\ref{fig:broadening} broadening, in Fig.~\ref{fig:sphericity} transverse sphericity and in Fig.~\ref{fig:thrust} thrust for the LHC and the Tevatron respectively. The differences in the histograms between LHC and Tevatron originate in the different jet definitions and $p_T$ thresholds. This leads to more spherical events and thus higher values for thrust and transverse sphericity at the Tevatron. For the backgrounds we include the processes that have the largest perturbatively calculable cross-sections. Most of these processes, in particular high-energy multijets and $W$+jets events, show a more pronounced pencil-like structure than the instanton events. Overall, analysing events with event shape observables provides a powerful method to discriminate instanton events from large Standard Model backgrounds.

%\begin{table}[]
%\centering
%\begin{tabular}{|l|l|l|}
%\hline
%Background process & Pythia  setting & No. of events   \\
%\hline
%Minimum bias & SoftQCD:nonDiffractive=on  & 200,000  \\
%\hline
% $qq\rightarrow X, qg\rightarrow X, gg\rightarrow X$ &HardQCD:All=on  & 100,000   \\
%(Hard QCD, low-energy) &PhaseSpace:pTHatMin = 5  &  \\
%\hline
% $qq\rightarrow X, qg\rightarrow X, gg\rightarrow X$ &HardQCD:All=on  & 100,000   \\
%(Hard QCD, high-energy) &PhaseSpace:pTHatMin = 100  &  \\
%\hline
%$W\rightarrow q\bar{q}+X$&WeakSingleBoson:ffbar2W = on&100,000\\
%\hline
%$t\bar{t}\rightarrow bq\bar{q} + \bar{b}q\bar{q} + X$ & Top:All=on & 100,000 \\
%\hline
%\end{tabular}
%\caption{A summary of the background processes we used and the associated Pythia settings.}
%\label{Table:backgrounds}
%\end{table}

\subsection{QCD Instanton Search at the LHC}

\subsubsection{Searches in high-luminosity LHC runs}
\label{sec:LHC}

%While under current and future high-luminosity runs the LHC provides very little sensitivity to QCD instantons

%early low-luminosity runs 
%open a novel path to search for QCD instanton events. 
As a result of the trigger cuts imposed, we find that the LHC has very little sensitivity to QCD instantons in current and future 
high-luminosity runs. QCD instanton events produce no isolated leptons or a large amount of missing transverse energy, and so appear only as multi-particle events consisting of soft jets. 

Missing transverse energy higher-level triggers require at least $E_{T\mathrm{mis}} \geq 70$ GeV while single jet triggers are as high as $p_{T,j} \geq 360$ GeV \citep{Aaboud:2016leb}. In Sec.~\ref{Sec:Recoil} we have shown that the emission of a hard jet from an initial state parton is not a viable strategy to produce an instanton. Further, the probability that one of the partons that originates in the instanton process has such a large momentum is very small as well. If one of the instanton-induced partons has a transverse momentum to pass the single-jet trigger requirements, the centre-of-mass energy of the instanton $\sqrt{s'}$ has to be at least of $\mathcal{O}(700)$ GeV. According to Table~\ref{Table:HadXSection}, this renders the hadronic instanton cross-section too small to be observable.

Thus, one would have to resort to multijet triggers, either with four jets of $p_{T,j} \geq 85$ GeV or six jets of $p_{T,j} \geq 45$ GeV. Both such trigger requirements result in for instantons fairly high partonic centre-of-mass energies of $\mathcal{O}(300)$ GeV.  Generating 100000 signal events as described in Sec.~\ref{sec:topology} and reconstructing them with the anti-kT jet algorithm, we find that none of the events passes multijet triggers, which results in an upper limit on the instanton cross-section that passes such trigger cuts of $\sigma_{pp\rightarrow I}^{\mathrm{trigger}} \lesssim 10$ fb. Disentangling instanton processes with less than $10$~fb of cross-section from large QCD backgrounds during the event reconstruction step is a highly challenging task. 

%Such high $E_{T}$ mean that we are working with incredibly high mass instantons where the cross-sections become very small and so it is not possible to observe instantons at the LHC, 

Due to increased pileup in future high-luminosity LHC runs and at future hadron colliders, e.g. the FCC-hh, trigger thresholds for jets will have to be increased, which will significantly reduce sensitivity to QCD instanton processes. Special trigger strategies would have to be developed for instantons to pass trigger requirements in such a jet-rich environment. One could speculate about the inclusion of event-shape observables directly in the trigger strategy and a highly optimised interplay between high-level and low-level triggers. As shown in Sec.~\ref{sec:topology} in Figs.~\ref{fig:leadpt}, \ref{fig:broadening}, \ref{fig:sphericity} and \ref{fig:thrust} instanton events have a very different event topology compared to QCD-induced multi-jet or resonance-associated production processes. Incorporating such observables in the trigger setup and reconstruction strategies might retain some sensitivity to instanton processes in future runs at high-energy hadron colliders.

\subsubsection{Search in low-luminosity LHC runs}

Rather than focusing on high-luminosity runs, we propose to pursue a different search strategy. The biggest obstacles to the discovery of QCD instanton processes are the high trigger thresholds, which are a necessity to avoid triggering on pileup in high-luminosity runs. Low-luminosity LHC runs had minimalistic trigger requirements \citep{ATLAS:2020cap}, i.e. min-bias triggers which required only a single charged track with an energy of 400 MeV. Remarkably, practically all QCD instanton events would pass min-bias triggers. ATLAS and CMS \cite{CMS:2018elu} both are in possession of un-prescaled min-bias datasets which are however often only used to determine the luminosity for low-pileup runs, rather than searching for new phenomena. 

To assess whether these datasets can provide sensitivity to QCD instanton processes, we generate event samples as outlined in Sec.~\ref{sec:topology} with a hadronic centre-of-mass energy of $\sqrt{s}=13$ TeV. For the event selection we require that each event should have at least six jets $n_j \geq 6$ with a minimum $p_{T,j} \geq 10$ GeV and that these jets form a thrust value of $\tau \geq 0.2$. This already confidently separates instanton signal events from QCD-induced background events. For an instanton with a minimum $\sqrt{s'} \geq 100$ GeV, which can be imposed through a requirement on the invariant mass of the final state jets, we find $\frac{s}{\sqrt{b}}=50.1$ and for $\sqrt{s'} \geq  200$ GeV we have $\frac{s}{\sqrt{b}}=7.1$. This shows a very good sensitivity for instanton processes in min-bias events, which can be further increased by lowering the $p_{T,j}$ requirements.

\bigskip
%%%%%%%%%%%%%%%%%%%%%%%%%%
\subsection{QCD Instanton Search at the Tevatron}
\label{sec:tevatron}
%%%%%%%%%%%%%%%%%%%%%%%%%%
\medskip

We deduce from the observations in Sec.~\ref{sec:LHC} that future runs at high-energy high-luminosity colliders are likely to become even less sensitive to QCD instanton processes. Consequently, looking into the other direction instead, e.g. at the Tevatron, might provide yet another way to search for QCD instantons. In the top row of Table \ref{Table:HadXSection} we show the hadronic cross-sections at Tevatron energies, depending on the partonic centre-of-mass energy of the instanton process.

We recast several jet-rich searches and measurements by CDF \citep{Aaltonen:2011vi,Aaltonen:2011kx,Collaboration:2012hz}. While a large fraction of instanton events would pass the trigger criteria, the event selection criteria applied in the analysis removed the predominant fraction of instanton events. Thus, the results provided in \citep{Aaltonen:2011vi,Aaltonen:2011kx,Collaboration:2012hz} did not allow to set an experimental constraint on the instanton cross-section. However, if this data was reanalysed and event reconstruction strategies following Secs.~\ref{sec:topology} and \ref{sec:LHC} were applied, the Tevatron could set stringent limits on the hadronic instanton cross-section.

%At the Tevatron, the $E_{T}$ requirements were much lower due to the reduced instantaneous luminosity and C.oM. energy. As in the LHC we have only jets in the final state but the analyses with low $E_{T}$ jet triggers had low sensitivity as the triggers were prescaled(\DM{unneccessary} again and so very few events actually made it into the final analysis.) However we found that analyses using a multijet trigger were far more sensitive as instanton events tend to have a greater number of particles in the final state. However for all of the papers we examined, the selection criteria used meant that sufficiently few instantons made it into the final analysis to be able to constrain the instanton cross-section. \citep{Aaltonen:2011vi,Aaltonen:2011kx,Collaboration:2012hz} are some papers where a multijet trigger was used; it is our suggestion that if new selection criteria were applied to the events on tape then it may be possible to distinguish between instantons and background.  In Figs. \ref{fig:leadpt}, \ref{fig:broadening}, \ref{fig:sphericity}, \ref{fig:thrust} we show plots comparing the distributions of the lead jet $p_{T}$, the broadening, the sphericity and the thrust of instantons and our background processes.(\DM{unneccessary?} These plots suggest that discriminating between signal and background using kinematic variables such as $p_{T}$ will be difficult and one should use some sort of shape variable such as thrust.)

\bigskip
\section{Conclusions}
\label{sec:conclusion}
Instantons are the best motivated, yet unobserved, non-perturbative effects predicted by the Standard Model. Being able to study instantons in scattering processes would provide a new window to the phenomenological exploration of the QCD vacuum and it would allow the tensioning of non-perturbative theoretical methods developed for gauge theories with data. 

In our calculation we used the optical theorem to calculate the total instanton cross-section from the elastic scattering amplitude by carrying out an integral over the instanton collective coordinates, and taking into account the hard-hard initial state interactions calculated in \citep{Mueller:1990qa}. The
inclusion of these interactions is essential as it provides a cut-off for the integral over the instanton scale size $\rho$ which otherwise diverges in the IR in any QCD-like theory when no explicit external scales (such as the scalar field VEVs, highly virtual momenta or high temperature) are present. 
This theoretical approach was first presented and applied recently in~\cite{Khoze:2019jta}. 
We improved on the results of~\cite{Khoze:2019jta} here by using a more robust integration method by directly computing 
 integrals over all instanton--anti-instanton collective coordinates that correspond to positive modes of the quadratic fluctuation operators in the instanton--anti-instanton background. 
This resulted in an increase to the instanton cross-section by approximately an order of magnitude, compared to the saddle-point approximation used previously. We then also calculated the mean number of gluons in the final state using a novel and more direct approach based on computing the expectation value of the instanton--anti-instanton interaction potential.

\medskip

Most of the earlier studies of QCD instanton-induced processes, prior to Ref.~\cite{Khoze:2019jta}, were specific to deep-inelastic 
scattering (DIS)~\cite{Balitsky:1992vs,Ringwald:1998ek,Ringwald:2003px}. In this case, it was the deep inelastic momentum scale $Q$ that
was essential for obtaining infrared safe instanton contributions in the DIS settings and at relatively low CoM energies.
The \Hone and \Zeus Collaborations have searched for QCD instantons at the \HERA collider~\cite{Adloff:2002ph, Chekanov:2003ww, H1:2016jnv}.
In the electroweak sector of the Standard Model, phenomenological consequences of similar non-perturbative processes were also studied in detail in the literature, including recent papers~\cite{Ringwald:2018gpv,Papaefstathiou:2019djz,Ellis:2016dgb}, and references therein. In particular, Ref.~\cite{Khoze:2020paj} relied on applying the theoretical formalism developed here and in \cite{Khoze:2019jta}, with the conclusion that {\it electroweak} instanton contributions at colliders are exponentially suppressed at all energies.  

\medskip

In this paper we have re-examined the phenomenology of QCD instanton contributions to high-energy scattering processes at hadron colliders. We showed that although the instanton cross-sections are very large in a hadron collider; surprisingly such colliders have little sensitivity to instantons due to the trigger criteria necessary to reduce the data rate. Although instantons produce many final state particles, the event is isotropic and the energy is divided between all particles resulting in few particles with large $p_{T}$, one of the principle trigger requirements in a hadron collider. The higher energy instantons which could potentially pass such triggers have a vanishingly small cross-section and would not be seen in sufficient numbers in the LHC to be distinguishable from the QCD background. However examination of data collected with a minimum bias trigger \citep{ATLAS:2020cap,CMS:2018elu} showed that it should be possible to either discover instantons or severely constrain their cross-section with such data, which was previously only used for luminosity calibration. We also examined data from the Tevatron and showed that certain triggers should have recorded many instanton events on tape but the selection criteria used in later analyses would render the analyses insensitive to instantons. With a new set of selection criteria this would also be another possible avenue for discovery. 

\bigskip
\section*{Acknowledgements}
We thank Deepak Kar, Frank Krauss and Matthias Schott for helpful discussions. We acknowledge funding from the STFC under grant ST/P001246/1.
%\clearpage
\vspace{1.0cm}
%\appendix
%\include{variables}

\startappendix
\Appendix{A. Instanton--anti-instanton valley configuration}
\label{sec:appA}
\medskip

The forward elastic scattering amplitude is obtained from the LSZ-reduced Green's function is calculated using the the path integral in the 
instanton--anti-instanton background,
\begin{equation}
G\left(p_{1},p_{2},p_{1},p_{2}\right)=\int  DA_\mu [Dq D\bar{q}]^{N_f} \, \prod_{i=1}^{4}A_{\mathrm{LSZ}}\left(p_{i}\right)e^{-S_{E}[A_\mu,q,\bar{q}]}.
\end{equation}
The definition and the meaning of the instanton--anti-instanton field configuration is provided by the valley method approach of Balitsky and Yung,
\citep{Balitsky:1986qn,Yung:1987zp} and the computation of the instanton cross-section using the optical theorem approach follows the approach
developed in \cite{Khoze:1990bm,Khoze:1991mx,Khoze:1991sa,Verbaarschot:1991sq}
and applied to QCD instantons at proton colliders in the recent paper~\cite{Khoze:2019jta}.

Usually when performing a functional integral such as this, we would expand the action around the minimum, recalling that the linear term vanishes as instantons satisfy the equations of motion, and we would get the functional determinant of $\frac{\delta^{2}S}{\delta A^{2}}$ but here we must be careful. If this operator possesses small or zero eigenvalues then the usual $(\det)^{-\frac{1}{2}}$ will become very large or singular as the Gaussian approximation fails. We must treat these zero/quasizero modes carefully. These modes arise when there is a symmetry or approximate symmetry of the system leaving the action unchanged. 

A typical example of a zero mode is the centre of the BPST instanton, the corresponding collective coordinate $x_0$ does not affect the value of the instanton
action and so translation is a symmetry. 
In general each symmetry of the system that is broken by the background field configuration (in our case the instanton) will have an associated collective coordinate, $\tau$, with zero mode $\frac{\partial A^{{\rm cl}(\tau)}}{\partial \tau}$,
where $A^{{\rm cl}(\tau)}$ denotes the background field.

Quasi-zero modes can be understood in a similar fashion even though they do not correspond to an exact symmetry of the system. A typical example of a 
quasi-zero mode is the separation between the positions of the instanton and the anti-instanton in the instanton--anti-instanton configuration.  At 
large separations, the individual (anti)-instantons interact very weakly and the collective coordinate that corresponds to their separation becomes a nearly flat direction of the instanton--anti-instanton action. Once again we denote the background instanton--anti-instanton field configuration
$A^{{\rm cl}(\tau)}$ and the quasi-zero mode is given by $\frac{\partial A^{{\rm cl}(\tau)}}{\partial \tau}$. In general $\tau$ will now
denote the set of all collective coordinates, for the zero and quasi-zero modes. 

The background field configuration with a quasi-zero mode (i.e. a nearly flat direction in the action parameterised by the $\tau$ coordinate)
can now be defined as a solution of the gradient flow equation, also known as the valley equation of Balitsky and Yung \citep{Balitsky:1986qn,Yung:1987zp},
\begin{equation}
\label{eq:valley}
\frac{\delta S}{\delta A}\bigg\rvert_{A=A^{{\rm cl}(\tau)}} \propto \, \epsilon^{2}\left(\tau\right) \frac{\partial A^{{\rm cl}(\tau)}}{\partial\tau}\,.
\end{equation}
If the background field is an exact classical solution, then the $\tau$-collective-coordinate parameterises an exact zero mode and we have
$\epsilon^{2}\left(\tau\right) = 0$ so the valley equation collapses to the Euler-Lagrange equation.
However, in the case of a quasi-zero mode, $\tau$ is a pseudo-flat direction; the action is not at the exact minimum at any fixed value of $\tau$.
In this case the equation \eqref{eq:valley_1} holds with a non-vanishing but small right hand side, so that $\epsilon^2 (\tau) \ll 1$.
The smallness of the parameter $\epsilon^2 (\tau)$ characterises how flat the corresponding quasi-zero mode is.

To proceed with our calculation of the Green's function one uses the Fadeev-Popov procedure \citep{Balitsky:1986qn,Yung:1987zp}:
\begin{equation*}
1=\int d\tau \left\lvert {\det}\left(\frac{d}{d\tau}\biggl\langle A-A^{{\rm cl}(\tau)},\frac{\partial A^{{\rm cl}(\tau)}}{\partial\tau}\biggr\rangle_{w}\right)\right\rvert
\, \delta\left(\biggl\langle A-A^{{\rm cl}(\tau)},\frac{\partial A^{{\rm cl}(\tau)}}{\partial\tau}\biggr\rangle_{w}\right)
\end{equation*}
\begin{equation}
= \int d\tau \,{\det}\left(\biggl\langle \frac{\partial A^{{\rm cl}(\tau)}}{\partial\tau},\frac{\partial A^{{\rm cl}(\tau)}}{\partial\tau}\biggr\rangle_{w}\right)
\, \delta\left(\biggl\langle A-A^{{\rm cl}(\tau)},\frac{\partial A^{{\rm cl}(\tau)}}{\partial\tau}\biggr\rangle_{w}\right)\,.
\label{eq:one}
\end{equation}
where $A^{{\rm cl}(\tau)}$ is the minimum of the action for fixed $\tau$ and
$\langle A,B\rangle_{w}$ denotes the scalar product or an overlap of two field configurations,
\begin{equation}
\langle A,B\rangle_{w} = \int d^4x \, w(x) A(x) B(x)\,.
\end{equation}
Note that the definition of the overlap above uses a positive weight function $w(x)$ -- the freedom to choose a convenient form of $w(x)$
is a well-known simplifying feature used in path integral expansions around instantons \cite{tHooft:1976snw,Levine:1978ge,Yung:1987zp} and will be utilised in what follows. Taking into account the weight factor, the valley equation reads,
\begin{equation}
\label{eq:valley_1}
\frac{\delta S}{\delta A}\bigg\rvert_{A=A^{{\rm cl}(\tau)}}  \,=\, \epsilon^{2}\left(\tau\right) w(x)\, \frac{\partial A^{{\rm cl}(\tau)}}{\partial\tau}\,.
\end{equation}

Inserting one of the factors of 1 in the form \eqref{eq:one} for each collective coordinate, and expanding the action $S(A)$ around
the background field $A^{{\rm cl}(\tau)}$,
\begin{eqnarray}
S(A) =\, S(A^{{\rm cl}(\tau)}) &+&
\bigr\langle\, \frac{\delta S(A^{{\rm cl}(\tau)})}{\delta A} , (A-A^{{\rm cl}(\tau)}) \bigr\rangle_w
\label{eq:Sexpan}\\
&+& \frac{1}{2}\bigl\langle(A-A^{{\rm cl}(\tau)}),
\Box(A^{{\rm cl}(\tau)})(A-A^{{\rm cl}(\tau)})\bigr\rangle_{w} \,+\, \ldots
\nonumber
\end{eqnarray}
 we get,
\begin{alignat}{1}
G=&N\int\prod_{i}d\tau_{i}\,
 {\det}\left(\bigl\langle \frac{\partial A_{\tau_{i}}}{\partial\tau},\frac{\partial A_{\tau_{j}}}{\partial\tau}\bigr\rangle_{w}\right)
\int DA \prod_{i}\delta\left(\bigl\langle A-A^{{\rm cl}(\tau)},\frac{\partial A^{{\rm cl}(\tau)}}{\partial\tau_{i}}\bigr\rangle_{w}\right)
\nonumber \\
&\qquad \quad \prod_{m=1}^{4}A_{\mathrm{LSZ}}\left(p_{m}\right)\, 
e^{
-S\left(A^{{\rm cl}(\tau)}\right)
-\frac{1}{2}\langle\left(A-A^{{\rm cl}(\tau)}\right),\Box(A^{{\rm cl}(\tau)})\left(A-A^{{\rm cl}(\tau)}\right)\rangle_{w}
}\,,
\label{eq:int_vall}
\end{alignat}

where $\Box\left(A^{{\rm cl}(\tau)}\right)=\frac{\delta^{2}S}{\delta A^{2}}\bigg\rvert_{A=A^{{\rm cl}(\tau)}}$. 
We note that the term linear in fluctuations in the expansion of the action (the second term on the right hand side
of \eqref{eq:Sexpan}) in fact does not contribute to the integral in \eqref{eq:int_vall}.
Indeed, the valley equation \eqref{eq:valley_1} requires that $\delta S/\delta A$ is proportional to $\partial A / \partial \tau$ when computed on 
our background configuration $A^{{\rm cl}(\tau)}$ and then the delta-function in the integrand  \eqref{eq:int_vall} ensures that this linear term vanishes.

Now we can perform the functional integration \citep{Yung:1987zp},
\begin{alignat}{1}
G=&N\int\prod_{i}d\tau_{i}\,\,
 {\det}\left(\bigl\langle \frac{\partial A_{\tau_{i}}}{\partial\tau},\frac{\partial A_{\tau_{j}}}{\partial\tau}\bigr\rangle_{w}\right)
 {\det}^{-{1}/{2}}\left(\bigl\langle\frac{\partial A_{\tau}}{\partial\tau_{i}},\Box^{-1}\left(A_{\tau}\right)\frac{\partial A_{\tau}}{\partial\tau_{j}}\bigr\rangle\right)
\nonumber \\
&\quad\qquad{\det}^{-{1}/{2}}\left(\Box\left(A_{\tau}\right)\right)\prod_{m=1}^{4}A_{\mathrm{LSZ}}\left(p_{m}\right)e^{ -S\left(A^{{\rm cl}(\tau)}\right)}.
\label{eq:A7}
\end{alignat}
Since the $\frac{\partial A^{{\rm cl}(\tau)}}{\partial\tau}$ play the role of zero- and quasi-zero modes of the action, they are the eigenfunctions of $\Box(A^{{\rm cl}(\tau)})$ and so,
\begin{equation}
\Box \bigl(A^{{\rm cl}(\tau)}\bigr)\,\frac{\partial A^{{\rm cl}(\tau)}}{\partial\tau_{i}}\,=\,\lambda_{i}\frac{\partial A^{{\rm cl}(\tau)}}{\partial\tau_{i}}\,.
\label{eq:qz_modes_1}
\end{equation}
This equation is valid at the leading order in the small parameter $\epsilon^2$ and follows from differentiating both sides of the valley equation 
with respect to $\tau$ and neglecting the $\epsilon^2(\tau)\, {\partial ^2A^{{\rm cl}(\tau)}}/{\partial\tau^2}$ term.

\noindent This allows us to simplify the product of the three determinants in \eqref{eq:A7} into
\begin{alignat}{1}
 {\det}^{1/2}  \left(\bigl\langle\frac{\partial A^{{\rm cl}(\tau)}}{\partial\tau_{i}},\frac{\partial A^{{\rm cl}(\tau)}}{\partial\tau_{j}}\bigr\rangle\right)
\, \left({\det}^{(2p)}\left(\Box\left(A^{{\rm cl}(\tau)}\right)\right)\right)^{-1/2}
\end{alignat}
 where ${\det}^{(2p)}$ denotes the determinant with the $2p$ zero and quasi-zero modes $\{\lambda_i\}_{i=1}^{2p}$ removed ($p$ modes for the
 instantons and $p$ modes for the anti-instanton). 
 
 To the leading order in the small-$\epsilon$ expansion we can also factorise the quadratic fluctuation determinant in the instanton--anti-instanton background
 $A^{{\rm cl}(\tau)}= A_{I\bar{I}}$ into the product of the instanton and the anti--instanton quadratic fluctuation determinants,
 ${\det}^{\left(2p\right)}\left(\Box\left(A_{I\bar{I}}\right)\right)\approx  {\det}^{\left(p\right)}\left(\Box\left(A_{I}\right)\right){\det}^{\left(p\right)}\left(\Box\left(A_{\bar{I}}\right)\right)$.
 
 \medskip
 This gives us finally \citep{Yung:1987zp},
\begin{equation}
G=\int d\mu_{1}d\mu_{2}\prod_{m=1}^{4}A_{LSZ}\left(p_{m}\right)e^{-S\left(A^{{\rm cl}(\tau)}\right)}
\end{equation}
where
\begin{equation}
d\mu_{a}=N\prod_{i=1}^{p}d\tau_{a,i}\, {\det}^{1/2}\left(\langle\frac{\partial A_{a}}{\partial\tau_{a,i}},\frac{\partial A_{a}}{\partial\tau_{a,j}}\rangle\right)\left({\det}^{\left(p\right)}\left(\Box\left(A_{a}\right)\right)\right)^{-1/2},
\end{equation}
are the instanton and anti-instanton collective coordinate integration measurse.

\bigskip

Having established the form of the collective coordinate integrals for the instanton-anti-instanton case, what is left for us to determine is the
instanton--anti-instanton configuration itself and in particular its action as the function of (anti)-instanton collective coordinates.

The instanton--anti-instanton valley trajectory $A_\mu^{I\bar{I}}$ was obtained in  Ref.~\citep{Yung:1987zp} by finding an exact
solution of the valley equation \eqref{eq:valley_1} for a particular choice of the weight function $w(x)$ by exploring conformal invariance
of the classical Yang-Mills action.
The action on this configuration was computed in \cite{Khoze:1991sa} and \cite{Khoze:1991mx,Verbaarschot:1991sq} and it takes the form,
\begin{eqnarray}
S_{I\bar{I}}(z) \,=\, \frac{16\pi^2}{g^2} \left(3\frac{6z^2-14}{(z-1/z)^2}\,-\, 17\,-\, 
3 \log(z) \left( \frac{(z-5/z)(z+1/z)^2}{(z-1/z)^3}-1\right)\right)\,,
 \label{eq:SzdefA}
\end{eqnarray}
where  the variable $z$ is the conformal ratio of the (anti)-instanton collective coordinates,
\begin{equation}
z\,=\, \frac{R^2+\rho^2+\bar{\rho}^2+\sqrt{(R^2+\rho^2+\bar{\rho}^2)^2-4\rho^2\bar{\rho}^2}}{2\rho\bar{\rho}}\,.
\label{eq:zdefA}
\end{equation}
$z$ plays the role of the single negative quasi-zero mode of the instanton--anti-instanton valley configuration.

In the limit of large separation between the instanton centres, $R/\rho,\, R/\bar{\rho} \to \infty$, the conformal ratio $z\to R^2/\rho\bar\rho \to \infty$, and 
instanton--anti-instanton action $S_{I\bar{I}}(z) $ becomes the sum of the indiidual instanton and anti-instanton actions,
\begin{equation}
\lim_{z\to \infty} S_{I\bar{I}}(z) \,=\, \frac{8\pi^2}{g^2} +  \frac{8\pi^2}{g^2} \,+\, {\cal O}(1/z^2)\,=\, \frac{16\pi^2}{g^2} \,,
\end{equation}
and 
\begin{equation}
A_{\mu}^{I\bar{I}}(x) \, \longrightarrow\, A_{\mu}^{I}(x-x_0) \,+\, A_{\mu}^{\bar{I}}(x-x_0-R)\,.
\end{equation}
In the opposite limit of a vanishing separation between the instanton centres,  $R/\rho,\, R/\bar{\rho} \to 0$,
the conformal ratio $z\to 1$ and the expression for the action ${\cal S}(z)$ goes to zero.
This is in agreement with the expectation that in this limit the instanton and the anti-instanton annihilate to the perturbative vacuum $A_\mu=0$.

\begin{figure}[]
\begin{center}
\hspace{-.4cm}
\includegraphics[width=0.5\textwidth]{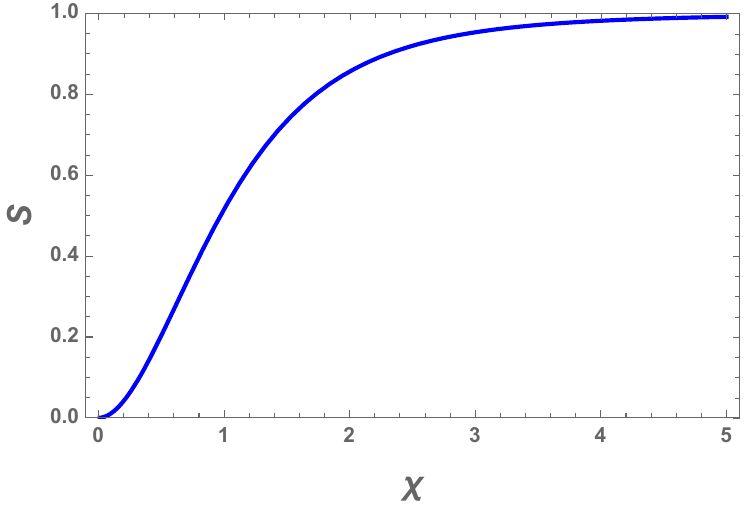}
\end{center}
\vskip-.4cm
\caption{
The action \eqref{eq:Szdef} of the instanton--anti-instanton configuration as the function of $\chi=R/\rho$ in units of ${16\pi^2}/{g^2}$.
$S_{I\bar{I}}$  approaches sum of the individual instanton actions
at $\chi\to \infty$ where the instanton interaction vanishes, and $S_{I\bar{I}}\to 0$ at $\chi\to 0$ where the
 instanton and the anti-instanton 
mutually annihilate. }
\label{fig:action}
\end{figure}

We can plot the action $S_{I\bar{I}}$ as the function of the separation between the instanton centres $R$ normalised by the instanton scale sizes.
For simplicity, if we assume that the sizes are equal, $\rho=\bar{\rho}$ we can write down the action $S_{I\bar{I}}$ as the function of the vartiable
$\chi=R/\rho$. It is plotted in Fig.~\ref{fig:action} in units of ${16\pi^2}/{g^2}$.

%%%%%%%%%%%%%%%%%%%%%%%%%%%%%%
%%%%%%%%%%%%%%%%%%%%%%%%%%%%%%%

\newpage
%\bigskip
\Appendix{B. Integration over the relative orientations}
\label{sec:appB}
\medskip

To be able to integrate over relative orientations in the internal $SU(3)$ space, 
we need to know the form of the instanton--anti-instanton action for arbitrary values of their relative orientation matrix $\Omega$.
However our exact valley configuration is only known for the maximally attractive channel, i.e. where the interaction potential 
$U_{\rm int} (z, \Omega)$
is maximised over the relative orientations for at each fixed value of $z$. 

What is known, however is the form of the interaction potential $U_{\rm int} (z, \Omega)$ in the limit of large separations.
In this large-separations regime (i.e. $z\gg1$) the instanton and the anti-instanton are known to have dipole-dipole interactions~\cite{Callan:1977gz},
\begin{eqnarray}
U_{\rm int}(z,\Omega) \,=\,  \frac{1}{z^2} \, \left(2\, {\rm tr} O\, {\rm tr} O^\dagger - {\rm tr} (OO^\dagger)\right)\,+\, {\cal O}\left(\frac{1}{z^4}\log z  \right) \,,
\end{eqnarray}
where $O$ is the $2\times2$ matrix in the upper-left corner of the $3\times 3$ matrix $\Omega$ describing the relative instanton--anti-instanton 
orientation.\footnote{The upper-left corner is selected by placing the instanton in the upper-left corner while allowing the 
anti-instanton to be anywhere in the of the $SU(3)$ internal space.}

Lacking the precise solution of the instanton--anti-instanton valley for general orientations at {\it arbitrary} separations, we will simply assume
that the full interaction potential can always be written in the form ({\it c.f.}~\eqref{eq:Uintinfz}),
\begin{eqnarray}
U_{\rm int}(z,\Omega) \,=\,  U_{\rm int}(z) \, \frac{1}{6} \left(2\, {\rm tr} O\, {\rm tr} O^\dagger - {\rm tr} (OO^\dagger)\right) \,,
\end{eqnarray}
where $U_{\rm int}(z)$ is the maximally-attractive-orienatation potential \eqref{eq:Uintdef},\eqref{eq:Szdef}. Clearly at large separations, to order $1/z^2$ this expression
coincides with the known dipole-dipole interaction.

We can now represent the integral over the relative orientations as follows,
\begin{eqnarray}
 \int d\Omega  \,e^{-\,\frac{4\pi}{\alpha_s(\mu_r)} \,{\cal S}(z, \Omega)} \,=\,
e^{-\,\frac{4\pi}{\alpha_s(\mu_r)}}\,  \int d\Omega  
\,e^{U_{\rm int}(z) \, \frac{1}{6} \left(2\, {\rm tr} O\, {\rm tr} O^\dagger - {\rm tr} (OO^\dagger)\right)
}
 \label{eq:op_Om2}
\end{eqnarray}

These type of integrals over $SU(3)$ matrices have been previously computed in the instanton literature, see Eq.~(2.15) in \cite{Balitsky:1992vs}:
\begin{eqnarray}
  \int d\Omega  
\,e^{\lambda \left(2\, {\rm tr} O\, {\rm tr} O^\dagger - {\rm tr} (OO^\dagger)\right)}\,=\,
\frac{1}{9\sqrt{\pi}}\, (2\lambda)^{-7/2} \,e^{6\lambda}
 \label{eq:op_Om3}
\end{eqnarray}
Substituting $\lambda = \frac{1}{6}\,U_{\rm int}(z)$ to the expression above, we now obtain the answer for our relative orientation integral in \eqref{eq:op_Om2},
\begin{eqnarray}
 \int d\Omega  \,e^{-\,\frac{4\pi}{\alpha_s(\mu_r)} \,{\cal S}(z, \Omega)} &=&
\frac{1}{9\sqrt{\pi}} \left(\frac{3}{U_{\rm int}(z)}\right)^{7/2} \,e^{-\,\frac{4\pi}{\alpha_s(\mu_r)} \,{\cal S}(z)} 
 \label{eq:op_Om_fin}
\end{eqnarray}
which agrees with the expression \eqref{eq:op_Om} quoted in section 2.

%\newpage
\bigskip

\bibliographystyle{inspire}
\bibliography{main}

\end{document}